\documentclass[useAMS,usenatbib]{mn2e}
\usepackage{amsmath}
\usepackage{amsfonts}%
\usepackage{amssymb}
\usepackage{graphicx, natbib, wasysym, subfiles, aas_macros}

\DeclareMathAlphabet{\mathbfsf}{\encodingdefault}{\sfdefault}{bx}{sl}

\usepackage{ulem}

\usepackage{color}

\title[Eccentricity of Massive Planets ]{Eccentricity Growth of Massive Planets  inside Cavities of Protoplanetary Discs}

\author[Romanova et al.]{\parbox{\textwidth}{M. M.~Romanova$^{1,2}$\thanks{E-mail of
corresponding author: \texttt{romanova@astro.cornell.edu}},
 A. V.~Koldoba$^{3}$, G. V.~Ustyugova$^{4}$,   D.~Lai$^{1,2}$, R. V. E.~Lovelace$^{1,2,5}$}
\vspace{0.4cm}\\
\parbox{\textwidth}{ 
$^{1}$Department of Astronomy, Cornell University, Ithaca, NY 14853-6801\\
$^{2}$Carl Sagan Institute, Cornell University, Ithaca, NY 14853-6801\\
$^{3}$Moscow Institute of Physics and Technology, Dolgoprudny, Moscow Region, 141700, Russia \\
$^{4}$Keldysh Institute for Applied Mathematics, Moscow, 125047,
Russia \\
$^{5}$Department of Applied and Engineering Physics, Cornell
University, Ithaca, NY 14853-6801}}
\date{\today}

\pagerange{\pageref{firstpage}--\pageref{lastpage}} \pubyear{2002}

\begin{document}
\label{firstpage}

\maketitle

\begin{abstract}

\noindent We carry out hydrodynamical simulations to study the eccentricity growth of a 1-30 Jupiter mass planet located inside the fixed cavity of a protoplanetary disc. The planet    
exchanges energy and angular momentum with the disc at resonant locations, and its eccentricity grows 
due to Lindblad resonances.   
We observe several phases of eccentricity growth where different eccentric Lindblad resonances dominate from 1:3 up to
3:5. 
The maximum values of eccentricity reached in our simulations are 0.65-0.75. 
We calculate the eccentricity growth rate for different planet masses and disc parameters 
and derive analytical dependencies on these parameters.
 We observe that the growth rate is proportional to both the planet's mass and the characteristic disc mass for a wide range of parameters. 
In a separate set of simulations, we derived the width
of the 1:3 Lindblad resonance.

\end{abstract}


\section{Introduction}

Many massive exoplanets have high eccentricities. There is a wide distribution of eccentricities at different planet masses and their distances 
from the star (see, e.g., Fig. 1 from \citealt{DebrasEtAl2021}, which is based on the recent  data from exoplanets.eu).
In cases of warm/cold Jupiters  (masses $0.5<M_p<5$ of Jupiter mass) approximately 50 per cent of planets
have   eccentricities $0.1<e_p<0.4$, while in cases of more massive planetary objects  ($5<M_p<50$),
the  eccentricities are even higher on average, with the eccentricity distribution almost uniform up to $e_p\approx 0.8$.

One class of mechanisms for eccentricity growth relies on the gravitational interaction
either through strong planet-planet scatterings (e.g., \citealt{RasioFord1996, LinIda1997,PapaloizouTerquem2001,ChatterjeeEtAl2008,JuricTremaine2008,MustillEtAl2017,AndersonEtAl2020,
LiEtAl2021}) or through secular perturbations from exterior stellar/planetary companions  (e.g., \citealt{HolmanEtAl1997,AndersonLai2017}). 
Another mechanism is the resonant interaction of a planet with an accretion disc.

A planet interacts with the disc due to the Lindblad and corotation resonances \citep{GoldreichTremaine1979,GoldreichTremaine1980}.
Lindblad resonances tend to increase the eccentricity of the planet, while corotation resonances suppress the eccentricity growth (see also \citealt{GoldreichSari2003,OgilvieLubow2003,TeyssandierOgilvie2016}). 
If a planet enters a low-density environment, the corotation torque becomes small, and eccentricity can grow due to the eccentric Lindblad resonances (ELRs)
(e.g., \citealt{ArtymowiczEtAl1991,dAngeloEtAl2006}). Such a situation appears if a massive planet clears a low-density gap in the disc, or 
if a planet  enters the low-density cavity surrounding a star.   
 A number of numerical simulations have been performed that show that eccentricity can increase due to the disc-planet resonant interaction
 (e.g., \citealt{PapaloizouEtAl2001,dAngeloEtAl2006,KleyDirksen2006,BitschEtAl2013,DunhillEtAl2013,RagusaEtAl2018,DebrasEtAl2021}).  However, only a small value of eccentricity has been obtained in most of the simulations, $e_p\sim 0.1- 0.25$ (e.g.,  \citealt{PapaloizouEtAl2001,dAngeloEtAl2006,KleyDirksen2006}).  
In many instances, the  authors concluded that eccentricity increases due to the 1:3 ELR (e.g., \citealt{KleyDirksen2006}). However, \citet{dAngeloEtAl2006} argued that the main resonances responsible for the eccentricity growth are the higher order ELRs: 2:4 and 3:5.

\citet{RagusaEtAl2018} performed very long simulations of planets in cavities of  low-mass discs. They observed regular patterns in which the disc and the planet exchange angular momentum. The maximum planet eccentricity in these simulations is $e_p\approx 0.12$.

 \citet{DebrasEtAl2021} developed a quasi-steady low-density 
cavity by taking a high viscosity 
in the cavity, and low viscosity in the disc
(in analogy with the dead disc ideas by \citealt{Gammie1996}). 
They investigated the migration of Jovian planets to the cavity and observed its eccentricity  growth. 
They showed that the planet's 
eccentricity can increase up to $e_p\approx  0.4$. They had to stop the  simulations because 
 a planet in eccentric orbit entered the region of wave damping placed around the inner boundary. 
 \citet{BaruteauEtAl2021} studied how a young planet shapes the gas and dust emission of its parent disc
using a post-processing radiative transfer model aimed at comparisons with ALMA observations.

\citet{RiceEtAl2008} used a different approach. They placed a planet into an empty cavity so that it interacted only with the disc and the star.
They observed the growth of eccentricity up to $e_p\approx 0.4$ in the case of very massive 20 Jupiter mass planets.
However, the authors considered  an unrealistically massive disc (to decrease computing time) and later scaled the results. \citet{TeyssandierOgilvie2016}
argued that scaling cannot be performed because the rate of exchange of the angular momentum between the disc and the planet 
 is not similar for discs of different masses. This issue of scalability has not been checked in numerical simulations.

In our earlier works, we experimented with low-density, high-temperature cavities that are in
pressure equilibrium with the  disc
  (e.g., \citealt{RomanovaEtAl2018}).  This approach (at low viscosity in the disc) provided a low-density
 cavity for a significant duration. We observed that the
eccentricity of a planet in the cavity grows due to resonant interaction with the disc. However, in many instances, a small amount of disc matter entered the cavity, causing  
 the planet's eccentricity to decrease due to co-orbital corotation torque.  In this work, to stop the matter from penetrating into the cavity, we fix the inner disc boundary. The low-density fixed-sized  cavity may be supported by various  physical  mechanisms, such as the magnetosphere of the star (e.g.,  \citealt{Konigl1991,Hartmann2000,RomanovaLovelace2006,RomanovaOwocki2015}), magnetic wind from the star (e.g., \citealt{LovelaceEtAl2008,SchnepfEtAl2015,Bai2016,WangGoodman2017}), or
evaporation of the inner disc due to UV radiation (e.g., \citealt{DullemondEtAl2007}).
By fixing the cavity border, we mimic such "real"  cavities. We also put zero density in the cavity because 
in many situations planet interaction with the cavity matter is expected to be less significant
 compared with the disc matter. In this approach, we do not need to calculate the gas flow inside the cavity.
 This approach is similar to that  
 used by \citet{RiceEtAl2008}. However, we take a lower-mass, more realistic disc.   
In this approach, we are able to observe different resonances responsible for eccentricity growth
and study the disc-planet interaction over long timescales. 
  
In our simulations, we observe that the eccentricity of the planet increases up to high values of  $e_p\approx 0.65-0.75$, which has never been obtained in earlier
studies. We perform simulations for different masses and different parameters of the disc and derived dependencies of
 the eccentricity growth rates on these parameters.

The plan of the paper is the following. In Sec. \ref{sec:theory}, we briefly review the main resonances. 
In Sec. \ref{sec:numerical-model}, we describe our numerical model and problem setup. In Sec. \ref{sec:refmodel}, 
we present the different phases in eccentricity growth and resonances  observed in simulations. 
In Sec. \ref{sec:dependencies}, we derive the dependence of eccentricity growth rate on planet mass and parameters of the disc.
We discuss different issues and applications in Sec. \ref{sec:discussion} and
conclude in Sec. \ref{sec:conclusions}. Appendix \ref{sec:numerical} presents the  details of our numerical model.   In Sec. \ref{sec:width}, we estimate the width of the 1:3 resonance.
Appendix \ref{sec:elliptical}, discusses possible effects of the ellipticity of the
orbit on positions of resonances.

\section{Resonances}
\label{sec:theory}

Eccentric resonances were studied  by \citet{GoldreichTremaine1978} (see also \citealt{GoldreichTremaine1979,GoldreichTremaine1980,Ward1986,Ward1997,Artymowicz1993a,Artymowicz1993b,Ogilvie2007,GoldreichSari2003,TeyssandierOgilvie2016}).
Following  \citet{TeyssandierOgilvie2016}, we summarize the properties of these eccentric resonances. 

The gravitational potential of a planet on the eccentric orbit can be expanded in a Fourier series :
\begin{equation}
\Psi(r,\phi,t)=\displaystyle\sum_{l,m}\psi_{l,m}(r)\rm{exp}[i({\it m}\phi-{\it l}\Omega_pt)].
\end{equation}
Here,  the coefficients $\psi_{l,m}\propto e_p^{|l-m|}$, where $e_p\leq 1$ is the planet's eccentricity.
For a planet in a circular orbit, $l=m$. For a planet in an eccentric orbit, at the first order in $e_p$, one
keeps terms with $l=m\pm 1$, where the plus/minus signs are relevant to the inner/outer resonances. Below, we consider only the outer resonances, and therefore we take  $l=m-1$.  

A planet interacts with the disc gravitationally and exchanges its energy and angular momentum. 
The rate of exchange is strongest at particular locations in the disc called resonances. Two types of eccentric resonances are important: Lindblad and corotation resonances.
The outer Eccentric Lindblad resonances (ELRs) correspond to locations in the disc where the perturbing frequency (in the rotating frame) matches the frequency of the disc:
$l\Omega_p-m\Omega=\Omega$. Taking $l=m-1$, one obtains  frequencies and radii in the disc corresponding to ELRs:
\begin{equation}
\frac{\Omega}{\Omega_p}=\frac{m-1}{m+1} , ~~~~ r_{\rm ELR}=a_p\bigg(\frac{m+1}{m-1}\bigg)^{2/3},  ~~~ m\geq 2~.
\end{equation}

 Eccentric corotation resonances (ECRs) are located in the disc where $l\Omega_p-m\Omega=0$. They occur at frequencies in the disc and  radii:

\begin{equation}
\frac{\Omega}{\Omega_p}=\frac{m-1}{m} , ~~~~   r_{\rm ECR}=a_p\bigg(\frac{m}{m-1}\bigg)^{2/3}, ~~~  m\geq 2~.
\end{equation}

There are also principal Lindblad resonances for a planet in a circular orbit (at $l=m$). The outer  Lindblad resonances (OLRs) are located 
at radii where
 $m\Omega_p-m\Omega=\Omega$ or: 

\begin{equation}
\frac{\Omega}{\Omega_p}=\frac{m}{m+1} , ~~~~   r_{\rm OLR}=a_p\bigg(\frac{m+1}{m}\bigg)^{2/3}, ~~~  m\geq 1~.
\end{equation}

The eccentricities ($e_p$ and $e_d$) of the planet and disc are coupled through resonant interactions.  Different resonances contribute to this process.  
According to \citet{Ogilvie2007} (see also \citealt{TeyssandierOgilvie2016}), a single 
ELR contributes to the evolution  of the eccentricities of the planet and the disc in the following way\footnote{Note that the authors 
used the complex eccentricity $E_p=e_p e^{-i\bar{w}}$, where $\bar w$ is the argument of pericentre.
Here, we neglected the precession of the orbit and take the absolute value $e_p=|E_p|$ .} 
 \begin{equation}
M_p a_p^2\Omega_p\bigg(\frac{\partial e_p}{\partial t}\bigg)_{\rm ELR}=\frac{GM_p^2}{M_\star}e_p {\cal B}^2\bigg(1 - \frac{{\cal A}e_d}{{\cal B} e_p}\bigg)\int{\Sigma }
F 2\pi r dr
\label{eq:dedt-p}
\end{equation}
 \begin{equation}
\Sigma r^2\Omega\bigg(\frac{\partial e_d}{\partial t}\bigg)_{\rm ELR}=- \frac{GM_p^2}{M_\star}{\Sigma{\cal A}{\cal B}e_p \bigg(1-\frac{{\cal A} e_d}{{\cal B}e_p}\bigg)} F .
\label{eq:dedt-d}
\end{equation}

The values of the coefficients  $\cal A$ and $\cal B$ and the resonant radii $r_{\rm res}$ are listed for small values of $m$ in Tab. \ref{tab:resonances}
 (see the full version of the table in \citealt{TeyssandierOgilvie2016}). 
Here 
\begin{equation}
F=w_L^{-1}\Delta\bigg( \frac{r-r_{\rm res}}{w_L}-1\bigg)
\end{equation}
is a function of the resonant radius $r_{\rm res}$, resonant width $w_L$, and dimensionless function $\Delta$, which describes the radial profile of the resonance.
  The width for 
 outer ELR has been estimated as (Teyssandier \& Ogilvie 2016, eq. 18):
\begin{equation}
{w_L}\approx r\bigg[\frac{h^2}{3(m+1)}\bigg]^{1/3}.
\label{eq:width}
\end{equation}

 Equations similar to \ref{eq:dedt-p} and \ref{eq:dedt-d} were derived for ECR where the relevant coefficients are $\cal C$ and $\cal D$.
Table \ref{tab:resonances} shows that higher-order resonances are located closer to the planet and the coefficients $\cal A-\cal D$ are larger.

\begin{table}
\begin{tabular}[]{ cc | ccc}
 \hline
 &&  Lindblad Resonances  &&    \\
\hline          
                 m & res & $r_{\rm res}/a_p$  &  $\cal A$    &  $\cal B$         \\
\hline          1  & 1:2& 1.587                 &               &                        \\
                 2 & 1:3 & 2.080                 & 0.607       & 1.849              \\ 
                 3 & 2:4 & 1.587                 & 5.201       & 3.594                  \\
                 4 & 3:5 & 1.406                 & 7.362       & 5.604                   \\
                5 & 4:6 & 1.310                 & 9.763       & 7.859               \\
\hline 
  &&  Corotation Resonances  &&   \\
\hline          
               m& res&  $r_{\rm res}/a_p$     &  $\cal C$      & $\cal D$    \\
\hline        2& 1:2 &   1.587                    &  1.723        & 0.620   \\ 
              3 & 2:3 &  1.310                    & 2.931        & 3.595    \\
              4 & 3:4 &  1.211                    & 4.111        & 4.751     \\
              5 & 4:5 &  1.160                    & 5.282        & 5.910    \\
\hline 

\end{tabular}
\caption{\textit{Top:} Values of $m$, type of resonance, ${\rm res}$, resonant radii $r_{\rm res}/a_p$ and coefficients  $\cal A$    and  $\cal B$ for ELRs.
\textit{Bottom:} same but for ECRs, where coefficients are $\cal C$    and  $\cal D$ .
  \label{tab:resonances}}
\end{table}

\section{Numerical Model}
\label{sec:numerical-model}

We consider  the orbital evolution of a massive planet (of Jupiter mass and higher) in an empty cavity surrounding a star.  
The planet  interacts 
gravitationally with the star and  accretion disc.

\begin{figure*}
     \centering
     \includegraphics[width=0.49\textwidth]{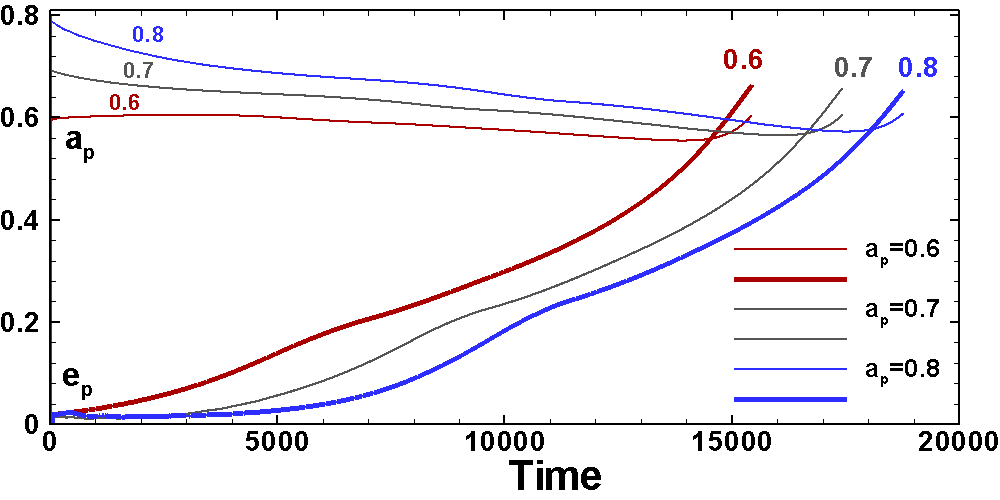} 
  \includegraphics[width=0.49\textwidth]{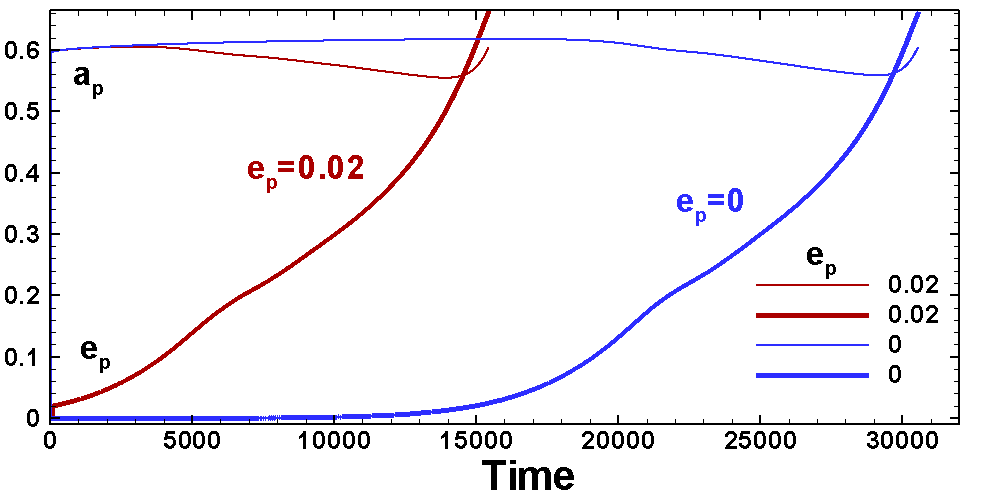} 
     \caption{\textit{Left panel:} Temporal evolution of the planetary orbit in models with different initial values of $a_p=0.6, 0.7, 08$ and initial eccentricity $e_p=0.02$.   
\textit{Right panel:} Evolution of the orbit with initial value of $a_p=0.6$, and eccentricity $e_p=0$ (blue curve) and $e_p=0.02$ (red curve).
\label{fig:orb-sax}}
\end{figure*}

\begin{figure*}
     \centering
     \includegraphics[width=0.8\textwidth]{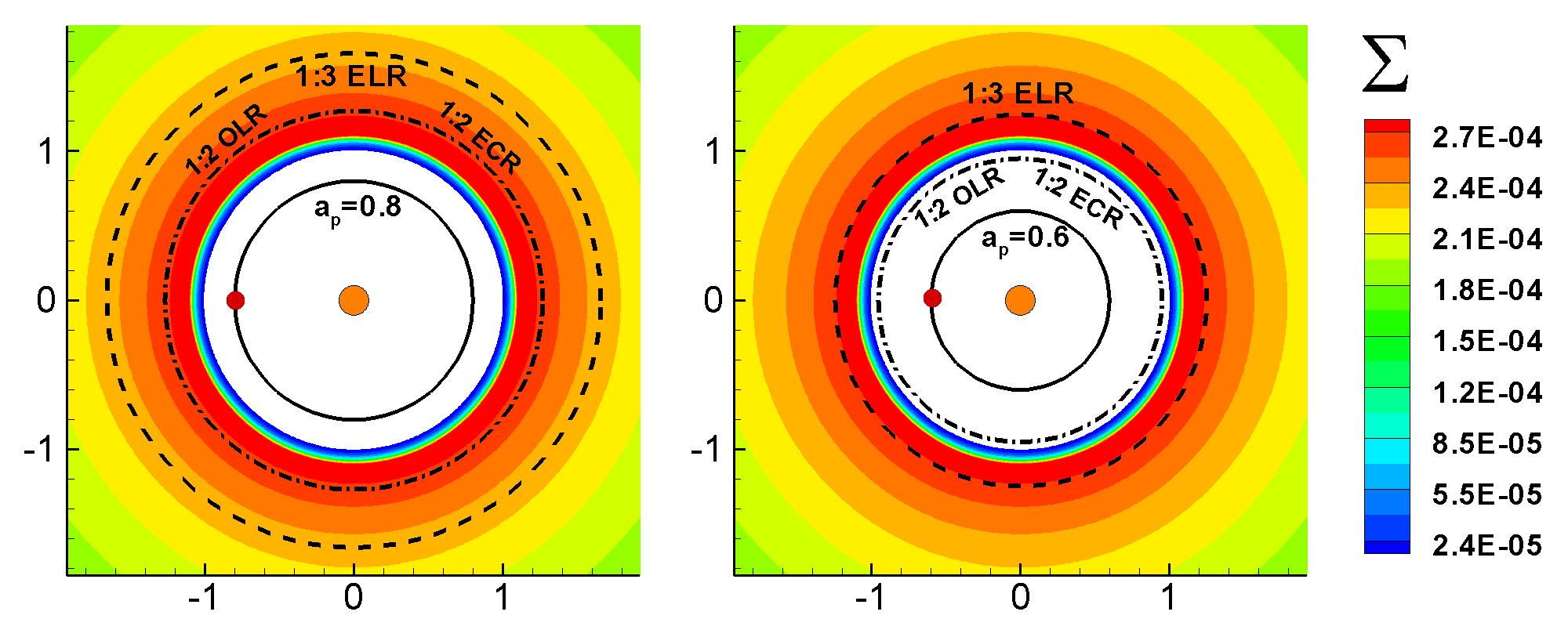} 
     \caption{\textit{Left panel:} Initial density distribution and location of resonances in  a model with an initial value of $a_p=0.8$. 
The solid line shows the position of the semimajor axis $a_p$, the dash-dot line shows the positions of the 1:2 OLR and 1:2 ECR, and the dashed line shows 
the position of the 1:3 ELR.  \textit{Right panel:} the same but for a planet with the initial value $a_p=0.6$.  
\label{fig:2d-res-2}}
\end{figure*}

\subsection{Initial and boundary conditions}
\label{sec:initial}

We place a  point-like star of mass $M_\star$ at the center of the coordinate system,  an empty cavity  
at radii $r<r_d$, and the disc  at radii $r_d<r<13.6r_d$.

We place a planet of mass $M_p$ in the cavity at a distance $a_p$ from the star. We take a slightly eccentric orbit
with $e_p=0.02$ (and $e_p=0$ in our test case).

We take a disc with an aspect ratio of $h=H/r=0.05$, which is determined at time $t=0$ at the inner edge of the disc,
$r=r_d$. 
Here, $H=(c_s/v_{\rm K}) r$ is
the scale height of the disc,  $c_s$ is the sound speed, and $v_K$ is the Keplerian
velocity. We investigate the dependence of our results on $h$ in Sec. \ref{sec:orb-thickness}.

 We take the power-law distribution of the surface density and pressure in the disc: 
\begin{equation} 
\Sigma(r) = \Sigma_d
     \left(\frac{r}{r_d}\right)^{-n} \quad
\Pi(r) = \Pi_d
     \left(\frac{r}{r_d}\right)^{-l} \quad. \label{eqn_pressure}
 \end{equation}
Here, $\Sigma_d$ and $\Pi_d$ are the
surface density and surface pressure at the radius $r=r_d$. 
We use values $n = l = 0.5$  (and investigate the dependence on $n$ in Sec. \ref{sec:orb-slope}).

We set an equilibrium distribution of  the 
azimuthal velocity $v_{\phi}$ in the disc by taking into account the 
balance of gravity and pressure gradient forces in the
radial direction:
\begin{equation}
v_\phi(r,z) = \sqrt{r \left(\frac{\partial \Phi}{\partial r} + \frac{1}{\Sigma} \frac{\partial \Pi}{\partial r} \right)}~,
\label{eq:vphi}
\end{equation}
where 
  $\Phi(r) = -GM_*/r$ is the
the gravitational potential of the star. 
This approach provides quasi-equilibrium
initial conditions in the disc.

We use ``free" boundary condition $\partial A/\partial r
=0$ for all variables $A$ at the inner (cavity) boundary and fixed boundary conditions at the outer boundary. 
 We use the procedure of
damping waves at the outer boundary, following \citet{Val-BorroEtAl2006}  (see their eq. 10). 
 Namely, we set the buffer zone for
damping at the outer part of the disc: $0.8 r_{\rm out}< r < r_{\rm
out}$. In addition, we put an exponential cut\footnote{Test simulations show that results are very close in models with and without an exponential cut. However, we keep an exponential cut for safety and also use it to model discs of different sizes  (see Sec. \ref{sec:orb-rexp}).}
 to the density and pressure distributions at $r=r_{\rm exp}$  to further decrease the possible influence of outer boundary conditions :
\begin{equation}
\Sigma(r)=\Sigma(r) e^{-\frac{(r-r_{\rm exp})}{\Delta}}, ~~\Pi(r)=\Pi(r) e^{-\frac{(r-r_{\rm exp})}{\Delta}},
\label{eq:rexp}
\end{equation} 
where
$r_{\rm exp} =0.5r_{\rm out}\approx 6.8r_d$ and $\Delta=0.22r_d$ in our Reference model. We take smaller values of $r_{\rm exp}$ and $\Delta$ in test models with the smaller-sized discs (see Sec. \ref{sec:orb-rexp}).

At the inner boundary, we place an exponential cut
 in the narrow region of $r_d<r<1.1r_d$ to have a smoother transition of density towards the inner edge of the cavity.  
We take $r_{\rm exp} =1.1 r_d$ and $\Delta=0.029 r_d$. 
We are not damping waves at the inner boundary.

\smallskip

We solve a full set of hydrodynamic equations in
 2D  including energy equation in entropy form.
We also  solve the equation for the planet's motion (see Sec. \ref{sec:numerical}).

We use a polar grid, which starts at the inner boundary $r=r_d$. 
The grid is centered on the star.  
It is evenly spaced in the azimuthal direction, where the number
of grid cells is $N_\phi=640$. In the radial direction, the size of the  grid cells progressively  increases 
such that the shape of grids is approximately square, and 
the number of grids is $N_r=308$. Test simulations were
performed using  finer and coarser grids. 
Simulations at finer grid  show convergence.
We chose the grid resolution $308\times 640$ in all simulation runs.

\begin{table}
\begin{tabular}[]{ c | c | c | c }
 \hline        $r_d$   (AU)                               &    $0.1$            &          $1$                                                 &  $10$     \\
 \hline
$q_d=M_d/M_0$                 &    $3\times 10^{-4}$     &    $3\times 10^{-4}$    & $3\times 10^{-4}$    \\
$\Sigma_d=q_d\Sigma_0$   [g cm$^{-2}$]               & $2.7 \times 10^5$                    & $2.7\times10^3$                   & 26.7                       \\
 $P_0$ [days]               & 11.6           & 365     & 11565  \\
\hline 
\end{tabular}
\caption{Reference dimensionless disc mass $q_d$, reference surface density  $\Sigma_d$, and time scale $P_0$  at different distances $r_d$.\label{tab:refval-real}}
\end{table}

\subsection{Dimensionalization}

The equations are written in dimensionless form and the results can be applied 
to cavities located at different distances from the star. We choose a reference scale,  $r_0=r_d$
and reference mass 
 $M_0 = M_*=M_\odot$.  The dimensionless mass of the planet is $q_p=M_p/M_*$.
For the convenience of presentation, we take  $M_*=M_\odot$ and measure the mass of the planet in
Jupiter masses. For example, 
in our reference model, we take $q_p=10^{-2}$ which is  $M_p=10$ in Jupiter masses.

The reference velocity is The Keplerian velocity  at $r=r_0$: $v_0 = \sqrt{GM_*/r_0}$.
We
measure time in the Keplerian period at $r=r_0$: $P_0 = 2
\pi r_0/v_0$. The reference 
surface density is $\Sigma_0 = M_0 / r_0^2$.
The
reference pressure is $\Pi_0 = \Sigma_0 v_0^2$.

\begin{figure*}
\centering
\includegraphics[width=0.9\textwidth]{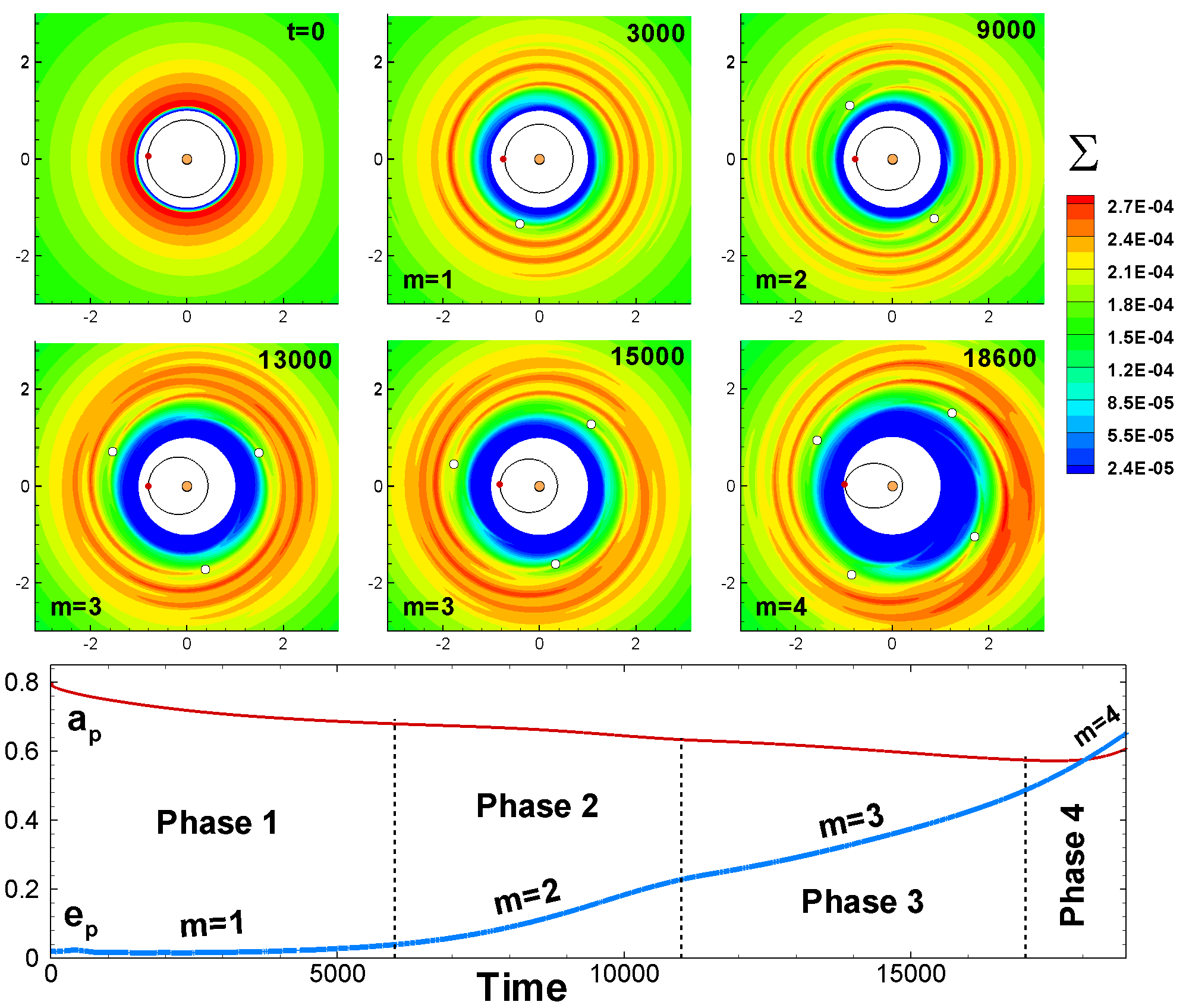}
\caption{Top 6 panels:
 waves forming in the inner disc in the Reference model  at different times. 
Time is measured in periods of rotation at $r_0=1$. The color background shows the surface density.
A small red dot shows the location of the planet, and small white circles show the locations of the spiral arms. 
The bottom panel shows the temporal evolution of the semi-major axis $a_p$ (red line) and eccentricity $e_p$ (blue line) of the planet's orbit.
The black vertical lines separate different phases of evolution,  in which  $m=1, 2, 3$ or $4$ spiral waves dominate.  
\label{fig:2d-r-4-6}}
\end{figure*}

 We also have the dimensionless parameter
in the code, $q_d$, which is used to vary the characteristic mass of the disc: $M_d=q_d M_0$. Thus, $q_d$ is the dimensionless 
characteristic mass of the disc: $q_d=M_d/M_0$. We vary $q_d$ in the range of  $q_d=10^{-2} - 10^{-4}$ and take $q_d=3\times 10^{-4}$ in the
reference model. 
We take into account that $M_0=\Sigma_0 r_0^2$ and obtain  $M_d=q_d \Sigma_0 r_0^2=\Sigma_d r_0^2$, where $\Sigma_d=q_d \Sigma_0$
is characteristic surface density in the disc. 
Therefore, $q_d$ also represents dimensionless characteristic surface density in the disc: $\bar\Sigma=q_d=\Sigma_d/\Sigma_0$.
Table \ref{tab:refval-real}  shows the reference 
dimensional values in models with $q_d=3\times 10^{-4}$ at  different $r_d$.

 We can compare values of reference surface densities in our model with 
values obtained for the Minimum-Mass Solar
Nebula (MMSN): 
  $\Sigma\approx 1700 (r/1\rm{AU})^{-3/2}$ g
cm$^{-2}$ \citep{Hayashi1981}.  
In our reference model, we take 
 $q_d=3\times 10^{-4}$, 
$r_0=1$ AU and obtain   
$\Sigma_{\rm d}=2.7\times 10^3$ g cm$^{-2}$. In the model with the lowest mass of the disc ($q_d=10^{-4}$)
we obtain
 $\Sigma_{\rm d}=8.9\times 10^2$ g
cm$^{-2}$. 
These values are close to those in MMSN.

\subsection{Reference model}

In the Reference model, we take a planet of mass $M_p=10$ and a disc with 
a dimensionless reference mass $q_d=3\times 10^{-4}$, thickness $h=0.05$, slope $n=0.5$, adiabatic index $\gamma=5/3$,
and viscosity coefficient $\alpha=3\times 10^{-4}$. 
We place a planet at the orbit with  eccentricity  $e_p=0.02$ and different semimajor axes: $a_p=0.6, 0.7$, and $0.8$ (see Tab.
 \ref{tab:ref-model} for parameters in Reference model).
The left panel of Fig. \ref{fig:orb-sax} shows the temporal evolution of $a_p$ and $e_p$  in models with different initial $a_p$. One can see that 
in all models $a_p$ decreases up to $a_p\approx  0.6$, and $e_p$ increases up  to $e_p\approx 0.65$. We note that in models
with initial values of $a_p=0.7$ and $0.8$, there is an initial interval of time with relatively fast inward migration
and slow growth of eccentricity. In contrast, in the model with initial $a_p=0.6$ inward migration is slow, while eccentricity increases
rapidly from the beginning of the simulation. 
 Analysis shows that in models with larger $a_p$ the principal 1:2 OLR is responsible for inward migration, while 
at $a_p=0.6$, this stage is absent, and the planetary eccentricity increases due to 1:3 ELR from the beginning of the simulation.
Fig. \ref{fig:2d-res-2} compares the positions of 1:2 OLR and 1:3 ELR (see Tab. \ref{tab:resonances} for positions of resonances) 
in models with  $a_p=0.8$ (left panel) 
and $a_p=0.6$ (right panel). One can see that in the model with $a_p=0.6$, the OLR is located inside the cavity, and this may be the reason, that
eccentricity increases due to 1:3 ELR from the beginning of simulations.

 In a test simulation run with $a_p=0.6$ and  $e_p=0$, we observe a long interval of time 
during which eccentricity increases very slowly (see the blue line in the right panel of Fig. \ref{fig:orb-sax}).   This is probably because the 1:2 ECR damps the eccentricity growth.
However, ECRs are saturated at small 
 values of $e_p$ (e.g., \citealt{GoldreichSari2003,OgilvieLubow2003}). We take $e_p=0.02$ in all our models.

\begin{table}
\begin{tabular}[]{ c  | c}
 \hline
Parameter   & Reference Model    \\
\hline          
                 reference disc mass                       & $q_d=3\times10^{-4}$                 \\
                reference surface density                 & ${\bar{\Sigma}}=3\times 10^{-4}$   \\
                initial semi-major axis                               & $a_p=0.6, 0.7, 0.8$                                 \\ 
                initial eccentricity                                      & $e_p=0.02$                                    \\
               coefficient of viscosity                               & $\alpha=3\times 10^{-4}$              \\      
           mass of the planet in stellar mass                    &    $q_p=10^{-2}$                   \\
              mass of the planet in Jupiter mass                &   $M_p=10$                          \\
                 semi-thickness of the disc                          & $h=0.05$                               \\
                slope in the density distribution                     & $n=0.5$                              \\
                 adiabatic index                                    & $\gamma=5/3$                       \\
\hline 
\end{tabular}
\caption{Parameters in the Reference model. 
  \label{tab:ref-model}}
\end{table}

\begin{figure*}
     \centering
     \includegraphics[height=0.35\textwidth]{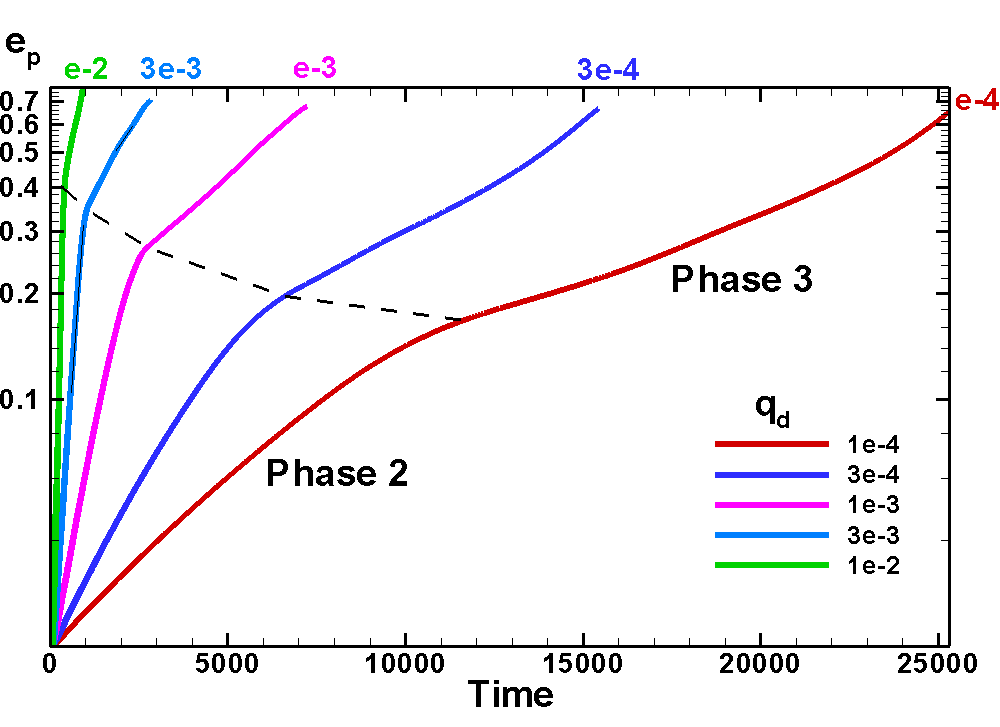} 
     \includegraphics[height=0.35\textwidth]{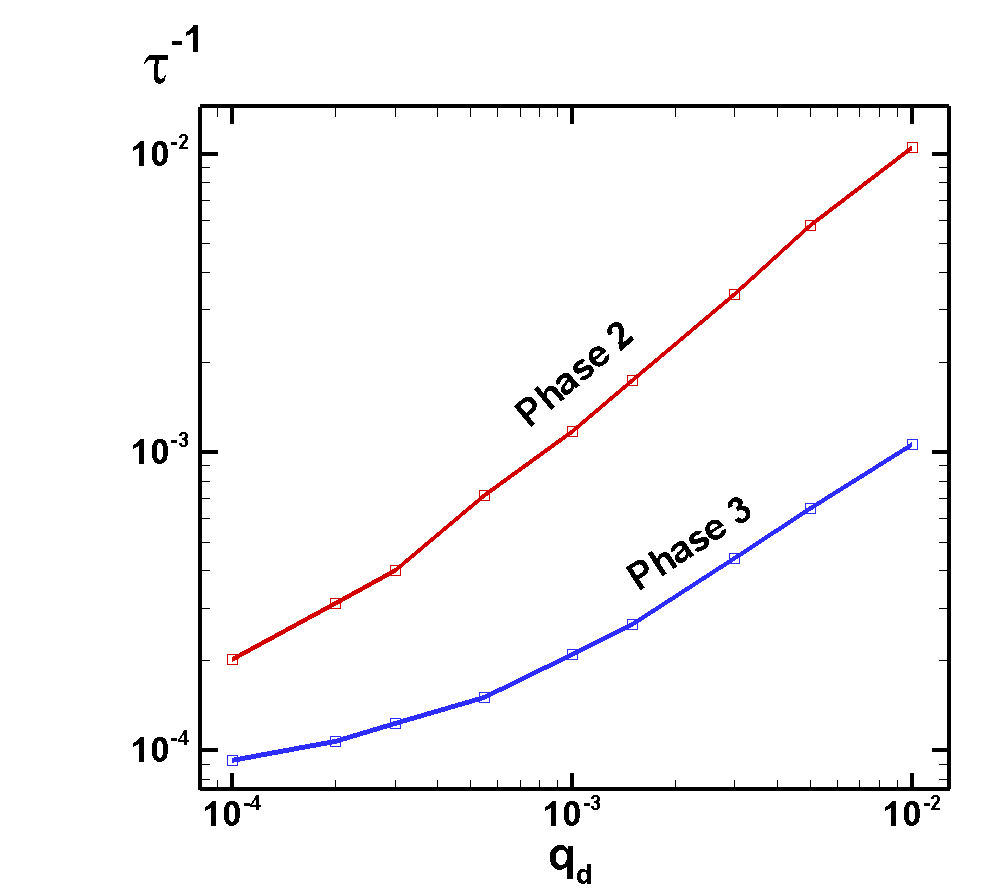} 
     \caption{\textit{Left panel:} Temporal variation of the eccentricity $e_p$ in models
with different reference disc mass $q_d$.   The dashed line separates Phases 2 and 3. 
 \textit{Right panel:}  Dependence of the eccentricity growth rate  $\tau^{-1}=(1/e_p) de_p/dt$ on $q_d$ 
at Phases 2 and 3.
     \label{fig:orb-dedt-sigma}}
\end{figure*}

\begin{figure*}
     \centering
     \includegraphics[width=0.8\textwidth]{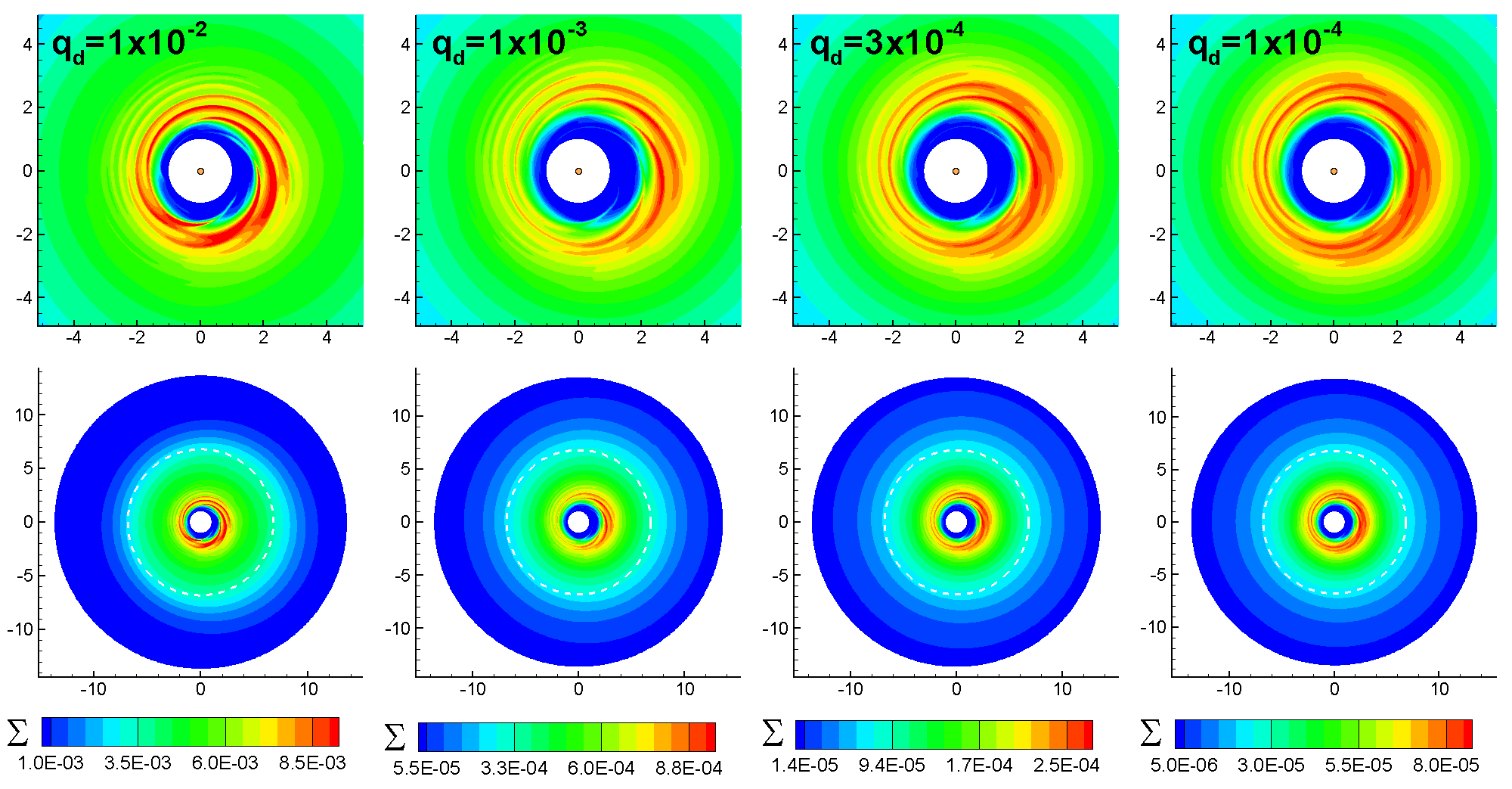} 
     \caption{\textit{Top panels:} Density distribution in models with different $q_d$ at the end of simulation run. The inner part of the simulation region is shown.
 \textit{Bottom panels:} Same, but in the whole simulation region. The dashed circle shows the radius of the exponential cut, $r=6.8$.
     \label{fig:2d-sigma-8}}
\end{figure*}

\begin{figure*}
     \centering
     \includegraphics[height=0.35\textwidth]{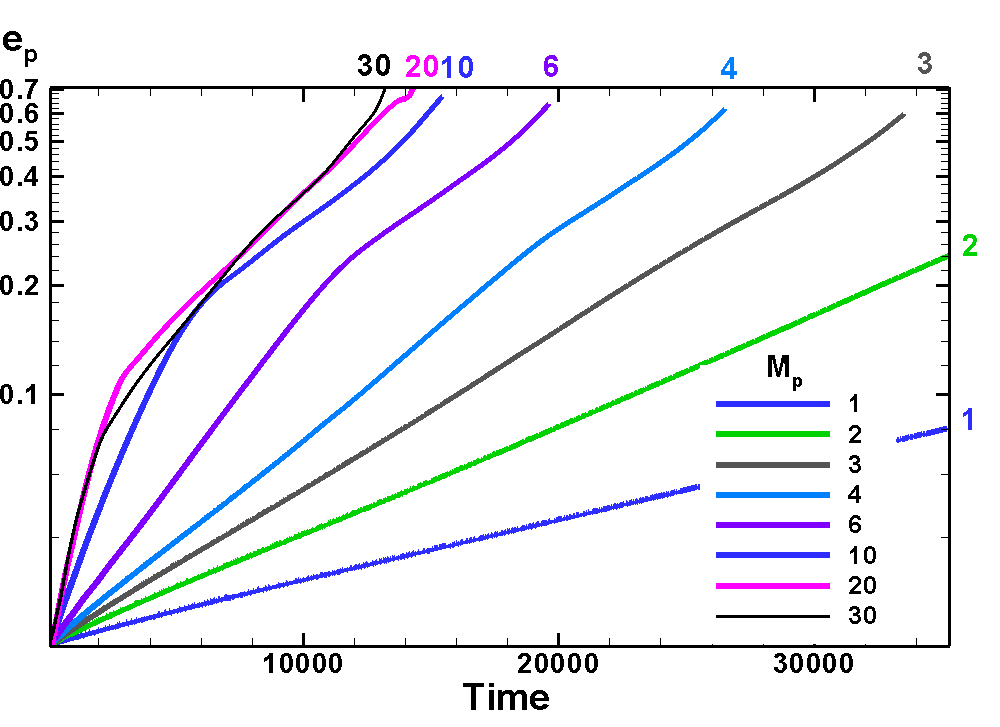} 
     \includegraphics[height=0.35\textwidth]{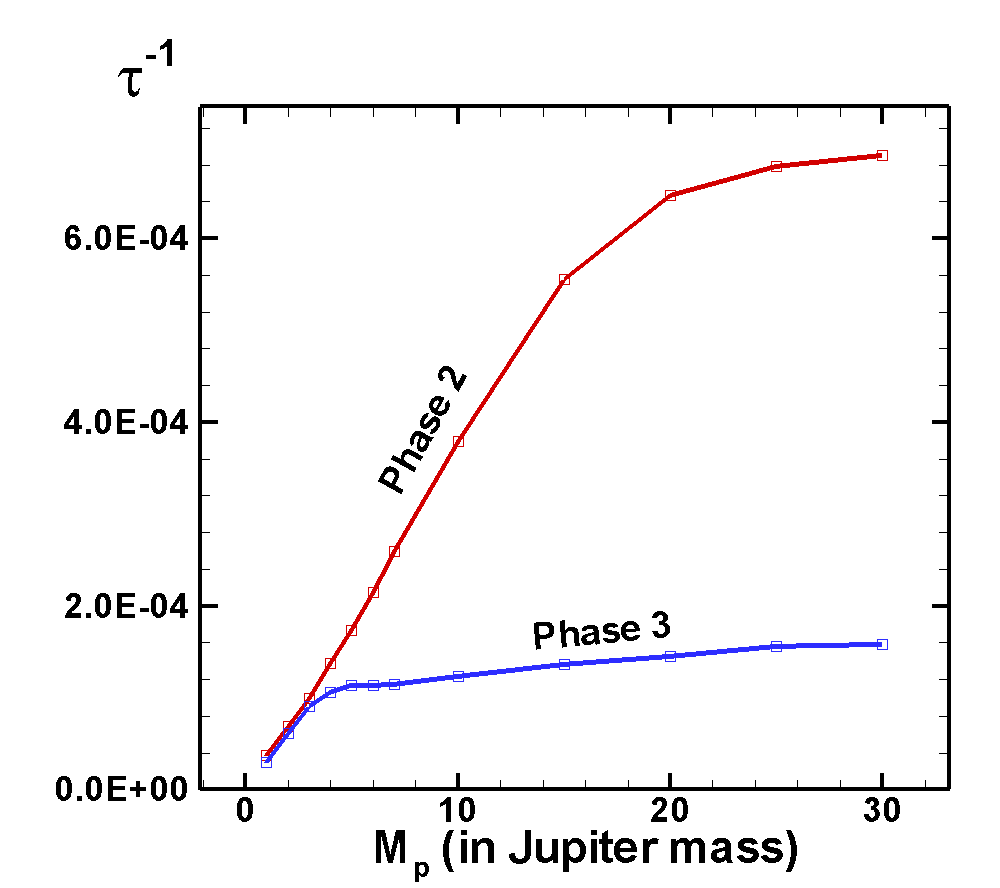} 
     \caption{\textit{Left panel:} Temporal variation of $e_p$ for different planet mass $M_p$ (in Jupiter mass)  in the Reference model.
 \textit{Right panel:} 
Dependence of the eccentricity growth  rate  $\tau^{-1}$ on the planet mass $M_p$.
     \label{fig:orb-dedt-mass}}
\end{figure*}

\begin{figure*}
     \centering
     \includegraphics[width=0.8\textwidth]{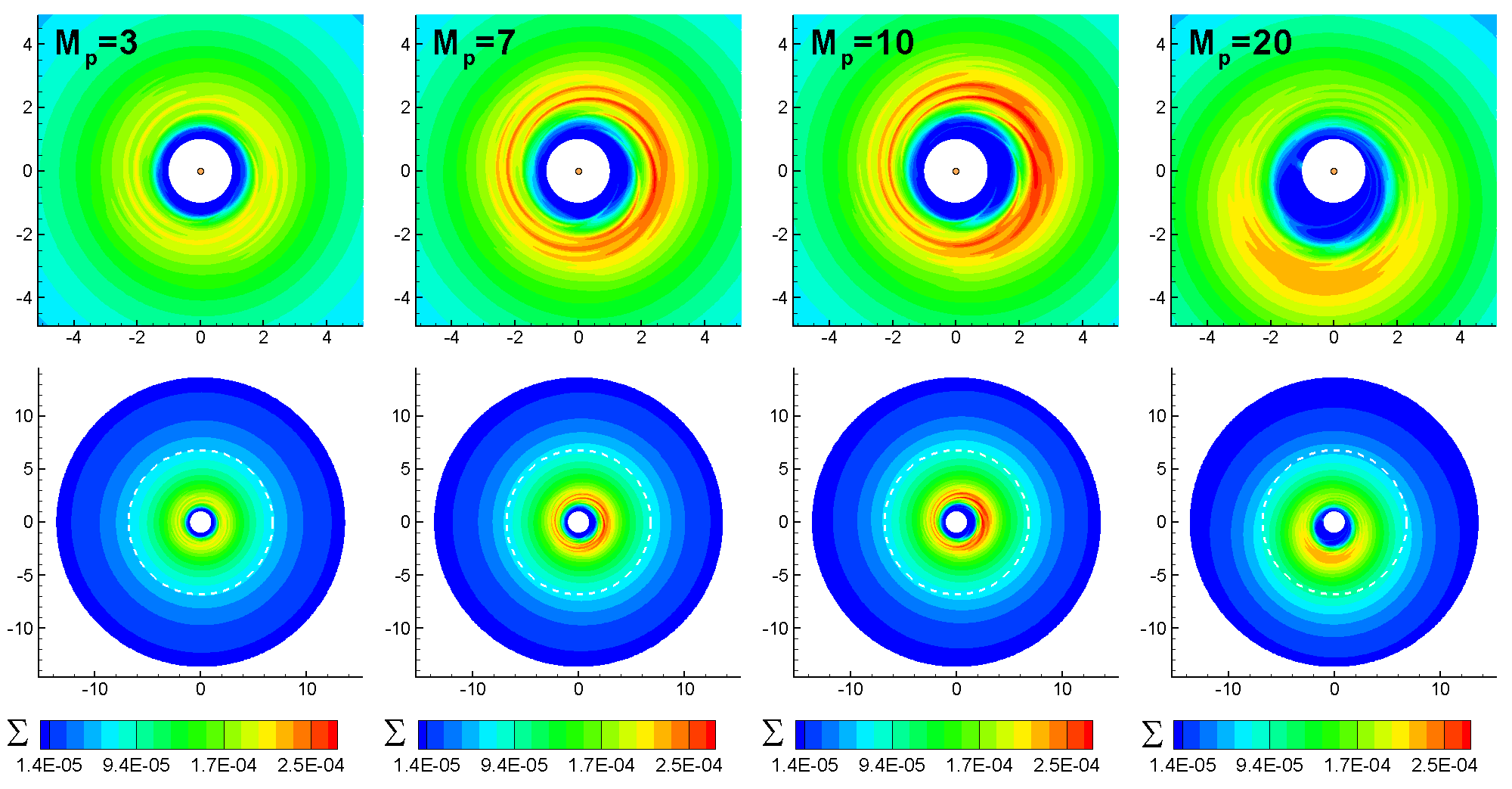} 
     \caption{\textit{Top panels:} Density distribution in models with different planet mass $M_p$ at the end of simulation run. The inner part of the simulation region is shown.
 \textit{Bottom panels:} Same, but in the whole simulation region.
     \label{fig:2d-mass-8}}
\end{figure*}

\begin{figure*}
     \centering
     \includegraphics[height=0.35\textwidth]{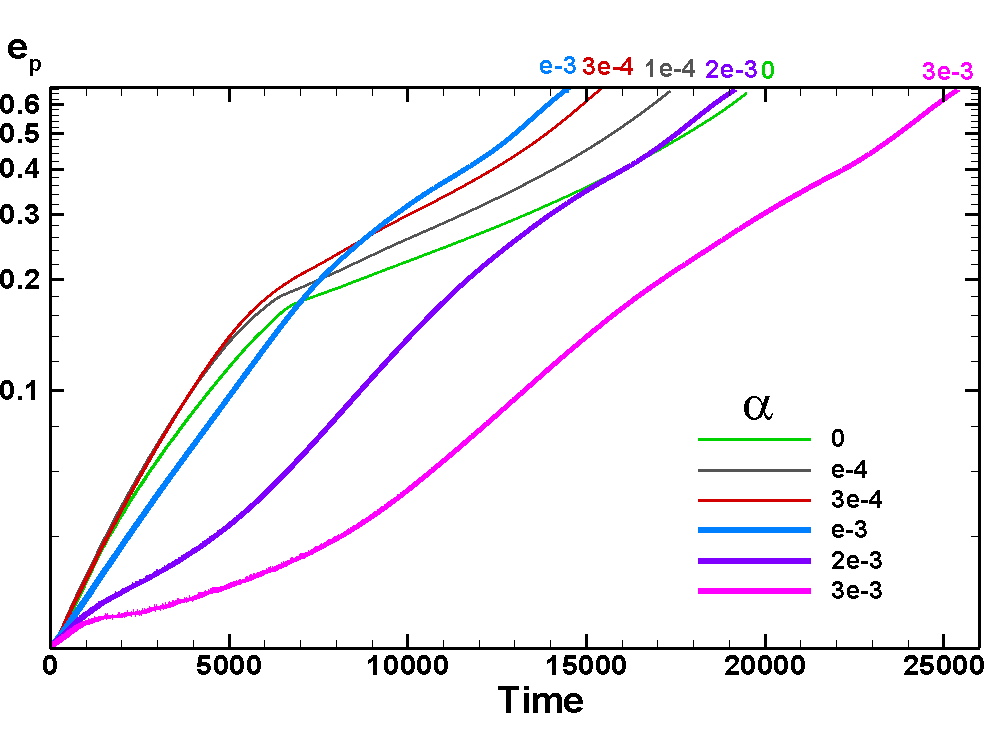}
     \includegraphics[height=0.35\textwidth]{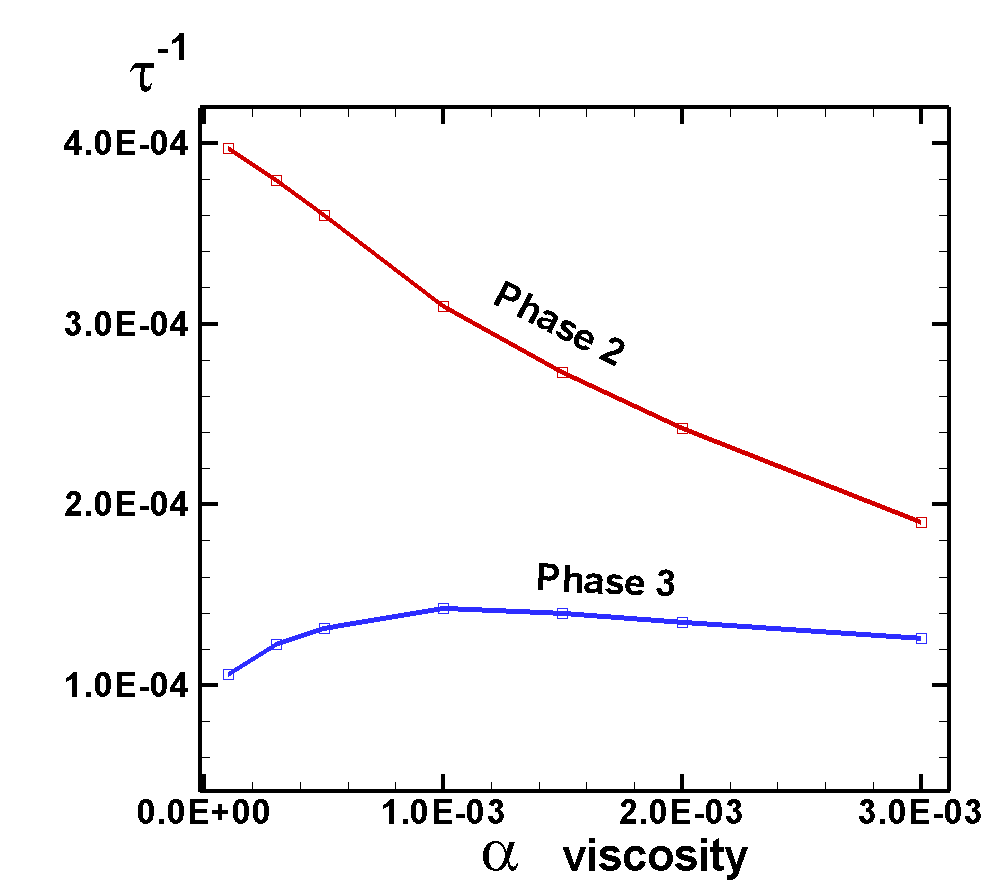} 
     \caption{\textit{Left panel:} 
Temporal variation of $e_p$ in models with different
 viscosity coefficients  $\alpha$.
\textit{Right panel:} 
Dependence of the eccentricity growth  rate  $\tau^{-1}$ on  $\alpha$.
     \label{fig:orb-dedt-visc}}
\end{figure*}

\begin{figure*}
     \centering
     \includegraphics[height=0.35\textwidth]{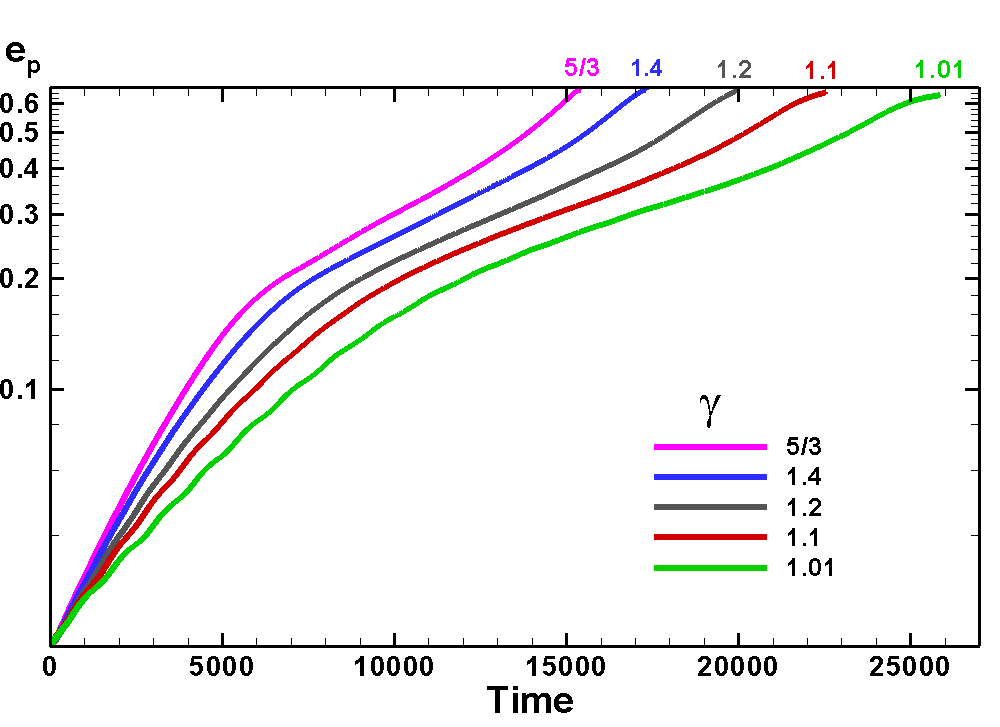} 
     \includegraphics[height=0.35\textwidth]{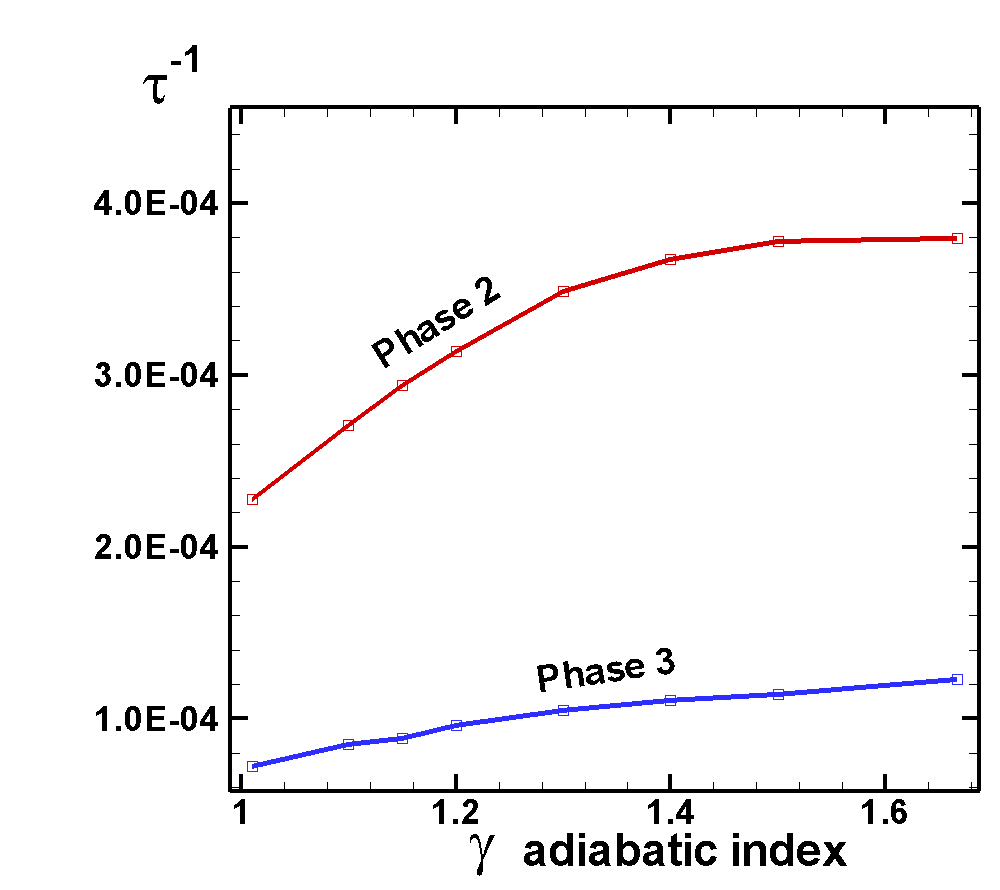} 
     \caption{\textit{Left panel:} Temporal variation of $e_p$ in models with different adiabatic index $\gamma$.
\textit{Right panel:} 
Dependence of the eccentricity growth  rate  $\tau^{-1}$ on $\gamma$.
     \label{fig:orb-dedt-gam}}
\end{figure*}

\begin{figure*}
     \centering
     \includegraphics[height=0.35\textwidth]{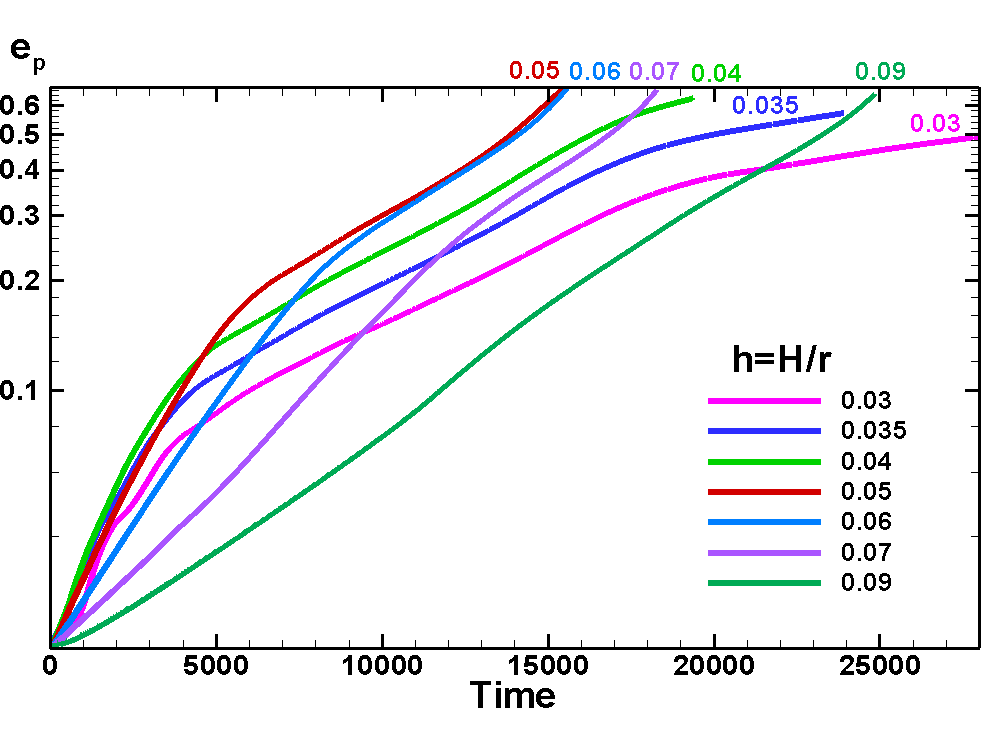} 
     \includegraphics[height=0.35\textwidth]{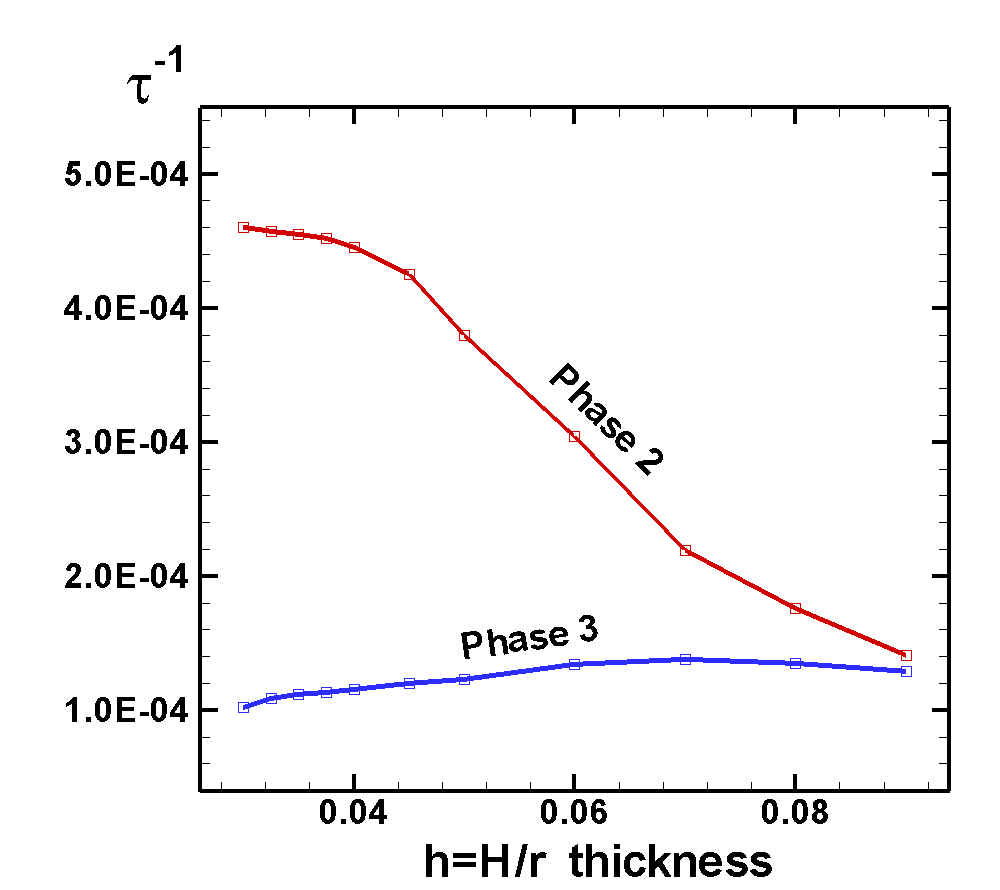} 
     \caption{\textit{Left panel:} Temporal variation of $e_p$ for different half-thickness of the disc, $h=H/r$. \textit{Right panel:} Dependence of the eccentricity growth rate $\tau^{-1}$ on $h$.
     \label{fig:orb-dedt-thick}}
\end{figure*}

\begin{figure*}
     \centering
     \includegraphics[height=0.35\textwidth]{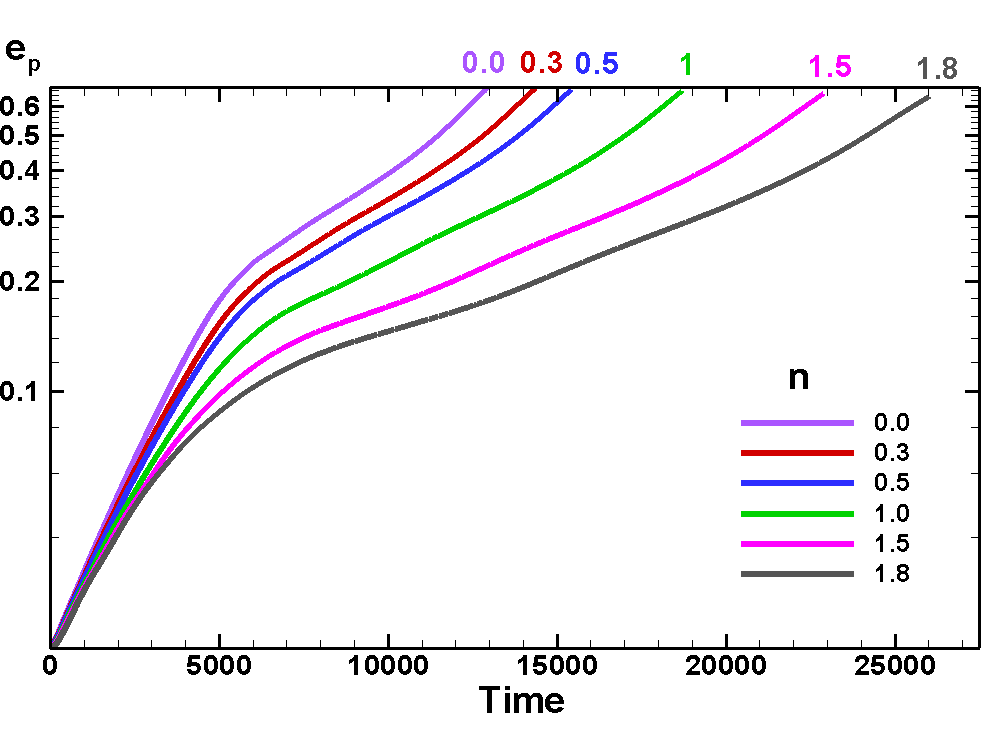} 
     \includegraphics[height=0.35\textwidth]{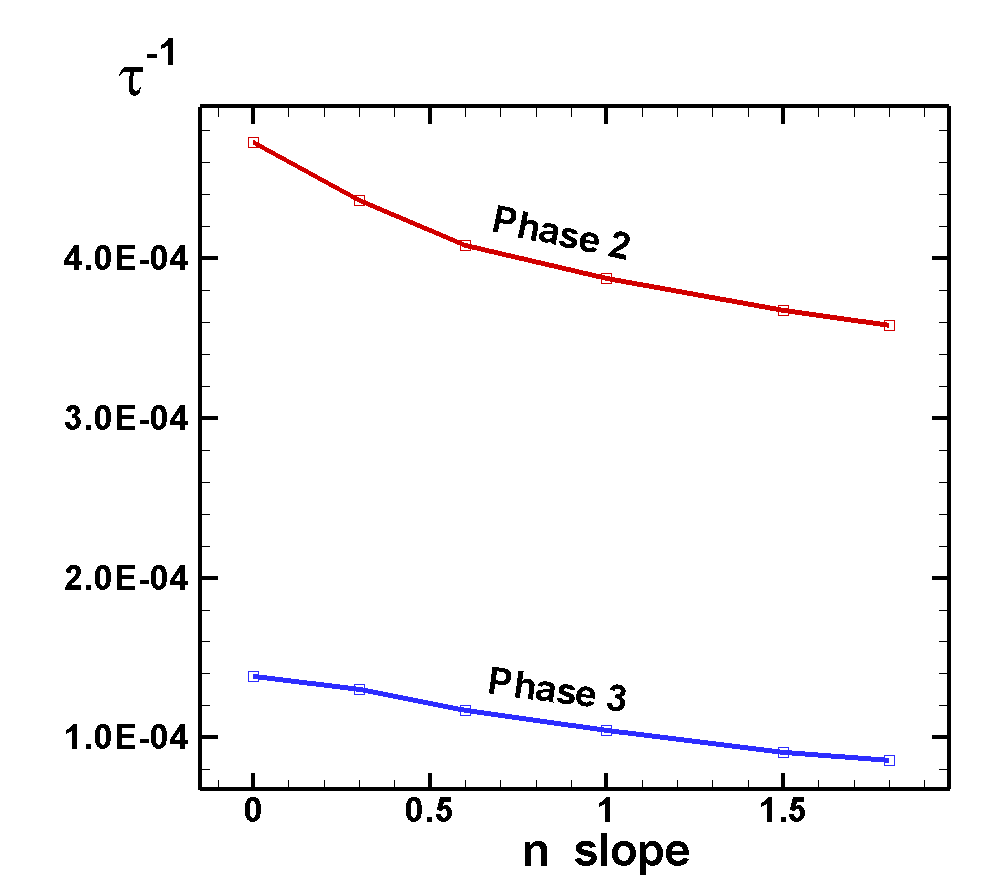} 
     \caption{\textit{Left panel:} 
 Temporal variation of $e_p$ in models with different slope indexes $n$: $\Sigma\sim r^{-n}$.
\textit{Right panel:} 
Dependence of the eccentricity growth rate $\tau^{-1}$ on $n$.
     \label{fig:orb-dedt-slope}}
\end{figure*}

\section{Phases of evolution} 
\label{sec:refmodel}

We observed several phases of orbit evolution, where different resonances can be responsible for eccentricity growth.
To demonstrate these phases,  we take a reference model with $a_p=0.8$.
Fig. \ref{fig:2d-r-4-6} shows the results of the simulations.
We observed that the planet excited spiral waves in the disc, but the number of arms  (the $m-$number) is different at different 
times during the orbit  evolution. The number of waves may signal the importance of a particular resonance. 
  Below, we describe these  phases in greater detail.   
 
\smallskip

\textbf{Phase 1} ($m=1$).  Initially, at times $0<t\lesssim 6000$, the planet excites a one-armed spiral wave in the disc  (see a  panel at $t=3,000$ 
  in Fig. \ref{fig:2d-r-4-6}). The bottom panel of the
same figure shows that, during this time interval, the semimajor axis decreases from 0.8 to 0.68 and eccentricity increases from 0.02 to 0.039.
This stage of evolution  can be associated with principal 1:2 OLR. At $a_p=0.8$, it is located at
$r_{\rm OLR,1:2} \approx 1.587 a_p\approx 1.27$  (see left panel of Fig. \ref{fig:2d-res-2}).
At the end of this phase, it is located at  the edge of the disc: $r_{\rm OLR,1:2} \approx 1.08$.
Corotation resonance 1:2 ECR is also located at the same distance.  However, at  $e_p=0.02$, it is saturated.

\smallskip

\textbf{Phase 2} ($m=2$). At $6000<t\lesssim 11000$, we observed  two-armed spiral waves in the disc
 (see an example at the panel at $t=9,000$ in Fig. \ref{fig:2d-r-4-6}). These waves are a sign of the 1:3 ELR.  
At this phase, $a_p$  decreases from 0.68 to 0.63, and $e_p$ increases from 0.039 to 0.23.  
At these values of $a_p$, this resonance is located at   
 $r_{\rm ELR,1:3} \approx 2.08 a_p\approx 1.41$ and $1.31$. Note that the 1:2 OLR is still in the disc: 
$r_{\rm OLR,1:2} \approx 1.587 a_p\approx 1.00 - 1.08$. However, the 1:3 resonance is stronger and dominates.

\smallskip

\textbf{Phase 3} ($m=3$). In the interval of time $11000\lesssim t\lesssim 17000$, we observed three-armed
spiral waves (see the panels at $t=13000$ and $t=15000$) which may be associated with the 2:4 ELR. During this phase, $a_p$ decreased from 0.63 to  0.57,
while $e_p$ increased from 0.23 up to  0.49.  At these (large) values of eccentricity, the  theoretical formulae can be applied only approximately.
In the linear theory of resonances, 2:4 ELR is expected to be at $r_{\rm ELR,2:4}\approx 1.587 a_p$, that
is at 1.00 and 0.90   at the beginning and the end of this phase, respectively.  At this phase, the 1:3 ELR is still in the disc, while the 2:4 ELR
is inside the gap. We suggest that several factors are important.  First, the resonance has a finite width and part of the 2:4 ELR can be located in the disc (see Sec. \ref{sec:width}).   
In addition,  the  planet has an elliptical orbit with the closest approach to the star in the pericentre, $a_{\rm per}=a_p (1-e_p)$, and furthest approach in the apocentre, $a_{\rm ap}=a_p (1+e_p)$. It spends most of the time in the  apocentre. If this factor 
is important, then the resonances will be in the disc  
(see Sec. \ref{sec:elliptical}).  The action of the 2:4 ELR is stronger than that of 1:3 ELR, probably because coefficients $\cal A$ 
and $\cal B$ are larger in case of 2:4 ELR (see Tab. \ref{tab:resonances}).

\smallskip

\textbf{Phase 4} ($m=4$) At $t>17000$, we observed 
another transition and
a final phase of evolution. Now, four-armed spiral waves were observed (see panel at $t=18600$). These waves may be associated with
  3:5 ELR. 
   During this phase, $a_p$ changes from 0.57 to 0.6, while eccentricity increases from $e_p=0.49$ to the final value of $e_p=0.66$. 
Using the formulae of linear theory\footnote{We should note that the linear theory has been developed for small values of eccentricity,
and therefore the positions of resonances may differ from those provided by the theory.} we obtain the positions of resonances:
 $r_{\rm ELR,3:5} = 1.406 a_p\approx 0.80$ and $0.84$, which are inside the gap.  Again, we suggest that the ellipticity of the orbit may increase 
the resonant radii (see Sec. \ref{sec:elliptical}).  Also,  the coefficients $\cal A$ 
and $\cal B$ are larger than in the lower-m resonances (see Tab. \ref{tab:resonances}). 

We stop simulations when a planet in apocentre reaches the disc-cavity boundary at $r=1$. We should note that the inner low-density region in the disc increases with time (see dark-blue areas in top panels of Fig. \ref{fig:2d-r-4-6}), and the eccentricity could increase to even higher values.

\section{Dependencies}
\label{sec:dependencies}

We varied the mass of the planet and the parameters of the disc and studied the dependence of the eccentricity growth
on different parameters.  We took, as a base, the model with $a_p=0.6$ and $e_p=0.02$ (thus skipping Phase 1).
We varied one parameter at a time.

\subsection{Dependence on the reference mass of the disc $q_d$ (reference surface density $\Sigma$)}
\label{sec:sigma}

We took several values of reference mass: $q_d=1\times 10^{-2},
3\times 10^{-3}, 10^{-3}, 3\times 10^{-4}, 10^{-4}$. 
The left panel of Fig. \ref{fig:orb-dedt-sigma}  shows that, 
in all models, the planet  eccentricity increases  to a high value.  In the model with $q_d=1\times 10^{-4}$,
it reached   $e_p\approx 0.65$, while 
 in the model with
$q_d=1\times 10^{-2}$,  it reached $e_p\approx 0.75$. 
We observed that eccentricity increases faster in models with more massive discs.
In all models, Phases 2 and 3 were observed, which are characterized by different slopes  and typical ``knees" between phases.

To calculate the rate of eccentricity growth $\tau^{-1}=(1/e_p)de_p/dt=d({\rm ln} e_p)/dt$, we plotted 
${\rm log}(e_p) $ versus time (like in the left panel of Fig. \ref{fig:orb-dedt-sigma}), then chose intervals of the linear (or almost linear) growth, and calculated      
the eccentricity growth rate $\tau^{-1}=d({\rm ln} e_p)/dt\approx \Delta({\rm ln} e_p)/\Delta t$ for each model. The right panel of Fig.  \ref{fig:orb-dedt-sigma} shows the dependence of
$\tau^{-1}$ on $q_d$ for Phases 2 and 3.  We observed that in Phase 2 at $q_d\gtrsim 3\times 10^{-4}$, the growth rate almost linearly varies with $q_d$ and can be presented as
$\tau^{-1}\sim({q_d}/{3\times 10^{-4}})^{0.95}$. 
In our model, the reference density in the disc $\bar\Sigma=q_d$, and therefore the dependence on $\bar\Sigma$ is
almost linear as well. This is in agreement with theoretical prediction:  $(1/e_p)de_p/dt\sim \Sigma$  (see, e.g., Eq. \ref{eq:dedt-p}).
This also means that our results can be scaled: simulations can be performed using high-density discs for which  the eccentricity growth rate
is high (and simulations are shorter). Subsequently, the results can be scaled to more realistic, lower-density discs  and longer time scales of 
eccentricity growth (e.g.,  in \citealt{RiceEtAl2008}).
 Fig. \ref{fig:2d-sigma-8} shows the density distribution in the disc for different $q_d$ at the end of the simulation run.  The top panels show that the resonant sets of waves are qualitatively similar despite the different densities of the disc.

In Phase 3, the dependence is more complex. In this case, we derive the dependence on $q_d$ in the vicinity of our reference value: 
 $\tau^{-1}\sim({q_d}/{3\times 10^{-4}})^{0.34}$.

\subsection{Dependence on the mass of the planet $M_p$}
\label{sec:mass}

We varied the planet
 mass from relatively small ($M_p=1$) to very large 
($M_p=30$) values.
The left panel of Fig. \ref{fig:orb-dedt-mass} shows that the planet's  eccentricity increases faster in models with larger $M_p$.
The right panel  shows that in Phase 2, the eccentricity growth rate   $\tau^{-1}$ systematically increases with $M_p$ up to
$M_p\approx 15$. However, the curve flattens at larger values of $M_p$. 
In Phase 3, the growth rate increases with $M_p$, but it is slower. 
The growth rates in the vicinity of our reference value of $M_p=10$ are:
$\tau^{-1}\sim({M_p}/{10})^{1.0}$ in Phase 2 and
 $\tau^{-1}\sim({M_p}/{10})^{0.23}$ in Phase 3.

 We should note that in the interval of masses  $1<M_p<15$  in Phase 2, the growth rate  $\tau^{-1}$ increases with a planetary mass almost linearly ($\tau^{-1}\sim({M_p}/{10})^{1.03}$) which is in accord with the theoretical prediction (see, e.g., Eq. \ref{eq:dedt-p}) \footnote{Note that the faster growth of eccentricity with the planet's mass has been also observed in simulations by \citet{PapaloizouEtAl2001}.}. 
Fig. \ref{fig:2d-mass-8} shows the density distribution in the disc for different $M_p$. The top panels show that at high mass $M_p$, the inner cavity becomes wider and non-axisymmetric. This may explain  the lower growth rate of eccentricity at high planet masses. The bottom panels of the same figure show the asymmetry of the whole disc 
increases at larger $M_p$.

\begin{figure*}
     \centering
     \includegraphics[height=0.45\textwidth]{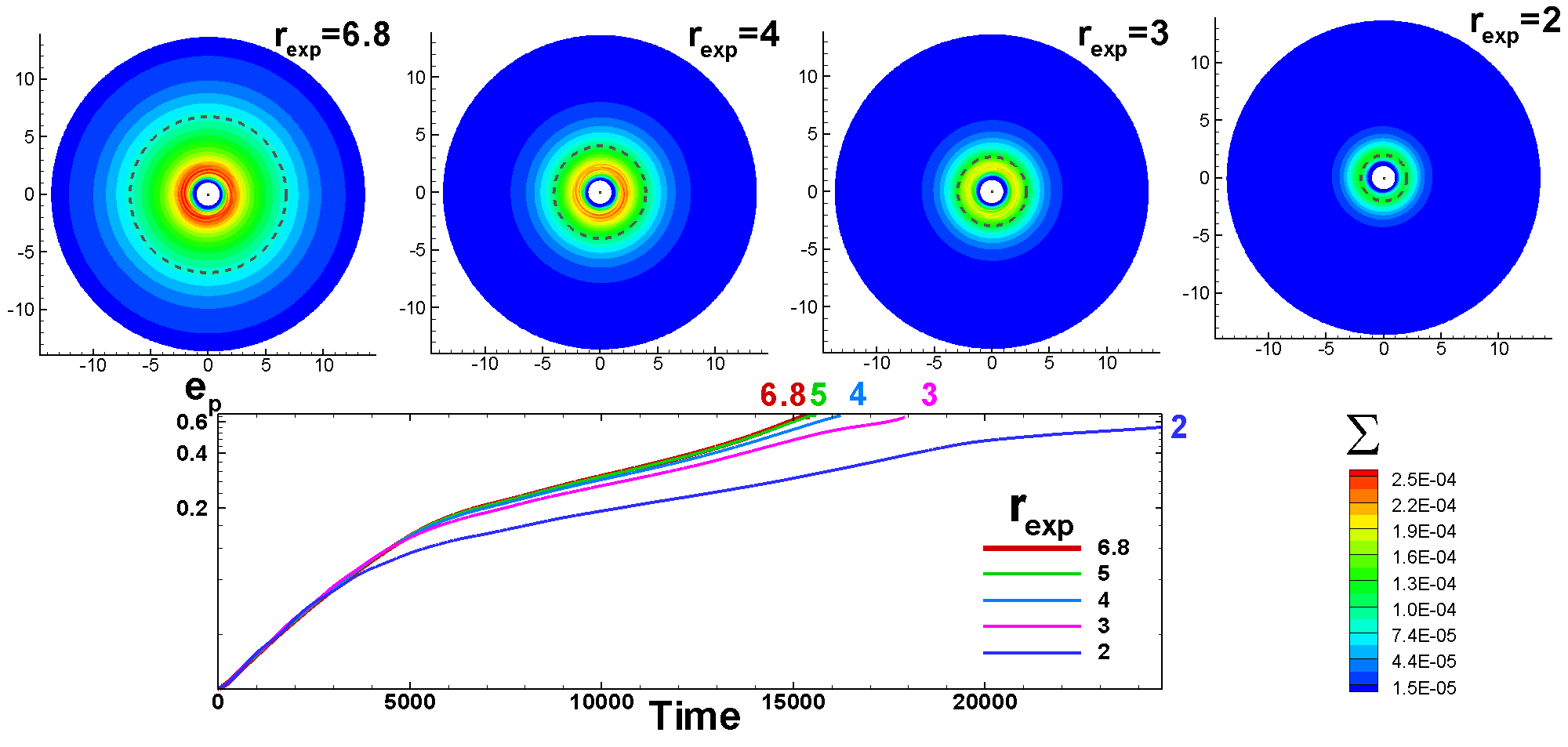} 
     \caption{
\textit{Top panels:} Density distribution at different exponential cuts $r_{\rm exp}$ at $t=10,000$. The radius of the exponential cut is shown as a dashed black line. \textit{Bottom panel:} Temporal variation of the eccentricity
in models with different $r_{\rm exp}$.
 \label{fig:2d-rexp-5}}
\end{figure*}

\subsection{Dependence on viscosity $\alpha$}
\label{sec:orb-visc}

We varied the viscosity coefficient $\alpha$ from  $\alpha=0$  to $\alpha=3\times 10^{-3}$. The left panel of Fig. \ref{fig:orb-dedt-visc} shows that in Phase 2 the eccentricity evolves slower with time for larger values of $\alpha$. The right panel shows that the eccentricity growth rate systematically decreases with $\alpha$ in Phase 2. We suggest that viscosity partly  damps waves excited at 1:3 resonance (see also \citealt{TeyssandierOgilvie2016}). The damping is weaker in Phase 3.
In the vicinity of our reference value of $\alpha=3\times 10^{-4}$, we derive dependencies
$\tau^{-1}\sim ({\alpha}/{3\times 10^{-4}})^{-0.11}$ for Phase 2
and $\tau^{-1}\sim({\alpha}/{3\times 10^{-4}})^{0.13} $ for Phase 3.

We see that our reference value of $\alpha=3\times 10^{-4}$ gives similar results as those at $\alpha=0$. Thus, the damping of waves is negligibly small.

\subsection{Dependence on the adiabatic index $\gamma$}
\label{sec:orb-gamma}

We varied the adiabatic index $\gamma$ from $\gamma=1.01$ to $\gamma=5/3$. The left panel of Fig. \ref{fig:orb-dedt-gam} shows that the planetary
eccentricity evolves slower in models with smaller values of $\gamma$. The right panel 
shows that in both phases, the eccentricity growth 
rate increases  systematically with $\gamma$ in the interval of $1.01<\gamma<1.3$ and slowly increases at larger values of $\gamma$. 
We solve an energy equation in the entropy form and use the equation of state in the form: $p= k\rho^\gamma$. Hence, we expect that the amplitude of resonant waves and the torque acting on the planet increases with $\gamma$. We see that the growth rate is approximately 1.8 times larger in models with $\gamma=5/3$ than in models with $\gamma=1.01$.  
 We estimate that at smaller values of $\gamma$ in the interval of $1.01\lesssim\gamma\lesssim 1.3$ the growth rate strongly increases with $\gamma$:  $\tau^{-1}\sim [{\gamma}/(5/3)]^{1.67}$ in Phase 2
and $\tau^{-1}\sim[{\gamma}/(5/3)]^{1.48}$ in Phase 3. However in the interval of   $1.4\lesssim\gamma\lesssim 1.66$, the dependence is not that strong:  $\tau^{-1}\sim [{\gamma}/(5/3)]^{0.18}$ in Phase 2
and $\tau^{-1}\sim[{\gamma}/(5/3)]^{0.59} $ in Phase 3. Currently, we do not know how to explain weaker dependence on $\gamma$ at larger $\gamma$ in Phase 2.

We should note that in many previous  simulations of planet migration in the cavity, the energy equation is not solved. Instead, a locally-isothermal equation of state is often adopted,  in which the temperature distribution depends only on the radius and is fixed with time (e.g., 
\citealt{PapaloizouEtAl2001,RiceEtAl2008,RagusaEtAl2018,DebrasEtAl2021}).
 Here, we cannot strictly  compare 
 models with and without the energy equation. 
However, we suggest that locally-isothermal models are closest to our models with 
 $\gamma=1.01$.

\subsection{Dependence on the thickness of the disc $h=H/r$}
\label{sec:orb-thickness}

We varied the  half-thickness of the disc $h$ from
$h=0.03$ up to $h=0.1$. The left panel of Fig. \ref{fig:orb-dedt-thick} shows the temporal evolution of
 eccentricity for different $h$.  
The right panel shows that in Phase 2, the eccentricity growth rate decreases when $h$ increases.  In Phase 3, the growth rate 
slowly increases with $h$.   In the vicinity of our reference value of $h=0.05$, we find dependencies
$ \tau^{-1}\sim ({h}/{0.05})^{-1.16}$
in Phase 2 and 
$ \tau^{-1}\sim ({h}/{0.05})^{0.37}$
in Phase 3.

A lower growth rate at larger $h$ in Phase 2 can be explained by the increase of viscosity with h: $\nu_{\rm vis}=\alpha c_s H$,  where both $H$ and $c_s$ increase with $h$. The viscosity damps the resonant waves (see also  Sec. \ref{sec:orb-visc}).

\subsection{Dependence on the density distribution in the disc (slope n)}
\label{sec:orb-slope}

Here,  we varied the slope $n$ in the initial density (and pressure) distributions: $\Sigma\sim r^{-n}$,
 $\Pi\sim r^{-n}$.
The left panel of Fig. \ref{fig:orb-dedt-slope} shows that
the orbit evolves more rapidly in models with flatter density distribution  (smaller values of $n$).  
The right panel of Fig. \ref{fig:orb-dedt-slope} shows that the eccentricity growth rate 
$\tau^{-1}$ 
decreases when $n$ increases.
We note that at steeper density distribution, there is less matter in the inner disc (where resonances operate) and the action of resonances is weaker. That is why $\tau^{-1}$ decreases with $n$ in both phases. 
 In the vicinity of our reference value, $n=0.5$, the dependence is 
$ \tau^{-1}\sim ({n}/{0.5})^{-0.10}$
in Phase 2 and 
$ \tau^{-1}\sim ({n}/{0.5})^{-0.15}$
in Phase 3.

\subsection{Analytical dependencies}

Here, we combine the dependencies derived in the above subsections for the planet's eccentricity growth rate:: 
\smallskip

\noindent In Phase 2:
\begin{eqnarray}
\nonumber {\tau_2}^{-1}&\approx& 3.8\times 10^{-4}\bigg(\frac{M_p}{10 M_J}\bigg)^{1.0}\bigg(\frac{q_d}{3\times 10^{-4}}\bigg)^{0.95}\bigg(\frac{h}{0.05}\bigg)^{-1.16}\\ 
               &\times&  \bigg(\frac{n}{0.5}\bigg)^{-0.1} \bigg(\frac{\alpha}{3\times 10^{-4}}\bigg)^{-0.1} \bigg(\frac{\gamma}{5/3}\bigg)^{0.18}
 \label{eq:phase2}
 \end{eqnarray}

\noindent In Phase 3: 
\begin{eqnarray}
 \nonumber\tau_3^{-1}&\approx& 1.23\times 10^{-4}\bigg(\frac{M_p}{10 M_J}\bigg)^{0.23}\bigg(\frac{q_d}{3\times 10^{-4}}\bigg)^{0.34}\bigg(\frac{h}{0.05}\bigg)^{0.37}\\
               &\times&  \bigg(\frac{n}{0.5}\bigg)^{-0.15} \bigg(\frac{\alpha}{3\times 10^{-4}}\bigg)^{0.13} \bigg(\frac{\gamma}{5/3}\bigg)^{0.59}
 \label{eq:phase3}
    \end{eqnarray}

We point out that these dependencies were derived in the vicinities of reference parameters. Away from these regions, the values of $\tau^{-1}$ should be taken from plots. For example, Fig. \ref{fig:orb-dedt-gam} shows that in Phase 2 the growth rate $\tau^{-1}$ increases with $\gamma$, as expected. However, in the vicinity of our reference value $\gamma=5/3$, it flattens. In several cases, however, the dependence is valid and can be used for a wide range of parameters (e.g.,  the  dependence on $q_d$ and  $M_p$).

\subsection{Dependence on the size of the disc}
\label{sec:orb-rexp}

We varied the size  of the disc by changing the radius of the exponential cut $r_{\rm exp}$ (and $\Delta$) in the initial conditions for the density and pressure distributions (see Eq. \ref{eq:rexp}). 
  In the reference model we have
$r_{\rm exp}=0.5 r_{\rm out}=6.8$. In test models we take
 $r_{\rm exp}=5, 4, 3$ and $2$ (and take smaller widths of the exponential transition $\Delta$). The top panels of Fig. \ref{fig:2d-rexp-5} show the density distribution in discs with different values of $r_{\rm exp}$.
The bottom panels of Fig. \ref{fig:2d-rexp-5} show that in models with $r_{\rm exp}=6.8, 5$ and $4$, the curves for the eccentricity
growth are  almost identical. 
 The eccentricity increases somewhat slower in the model with the smaller-sized disc, where $r_{\rm exp}=3$ (it increases  to the
maximum value during 17,900 rotations vs 15,300 in the reference model). The eccentricity  increases even slower
in the model  with the smallest disc, where $r_{\rm exp}=2$ (during 29,000 rotations). We conclude that resonant excitation of eccentricity occurs
when the disc is large enough to incorporate the width of the ELR.  
When the disc is smaller than ELR width (the model with $r_{\rm exp}=2$), then only  part of the resonance
is in the disc, and the action of the resonance is weaker. We conclude that the disc with the size of   $r_{\rm exp}=3$ 
would be sufficient for resonant excitation of eccentricity.


The total dimensionless mass of the disc is calculated by integrating the density distribution from  $r=1$ to $r=r_{\rm exp}$: 
\begin{equation}
{M}_{\rm d,tot}=\int_1^{r_{\rm exp}}{{\bar{\Sigma}} r^{-0.5} 2\pi r dr} = {\bar{\Sigma}} (4/3)\pi r^{1.5}\bigg|_1^{r_{\rm exp}}  .
\end{equation}
Taking our reference value of ${\bar\Sigma}=3\times 10^{-4}$, we obtain    ${M}_{\rm d,tot}\approx 2.1\times10^{-2}$ in our reference model ($r_{\rm exp}=6.8$),
 and 
${M}_{\rm d,tot}\approx 5.3\times10^{-3}$ in the model with $r_{\rm exp}=3$.
The disc mass is approximately twice as larger as the planet mass $q_p=10^{-2}$ in our reference model and twice as smaller
as the planet's mass in the model of a small disc.

\begin{figure}
     \centering
     \includegraphics[width=0.5\textwidth]{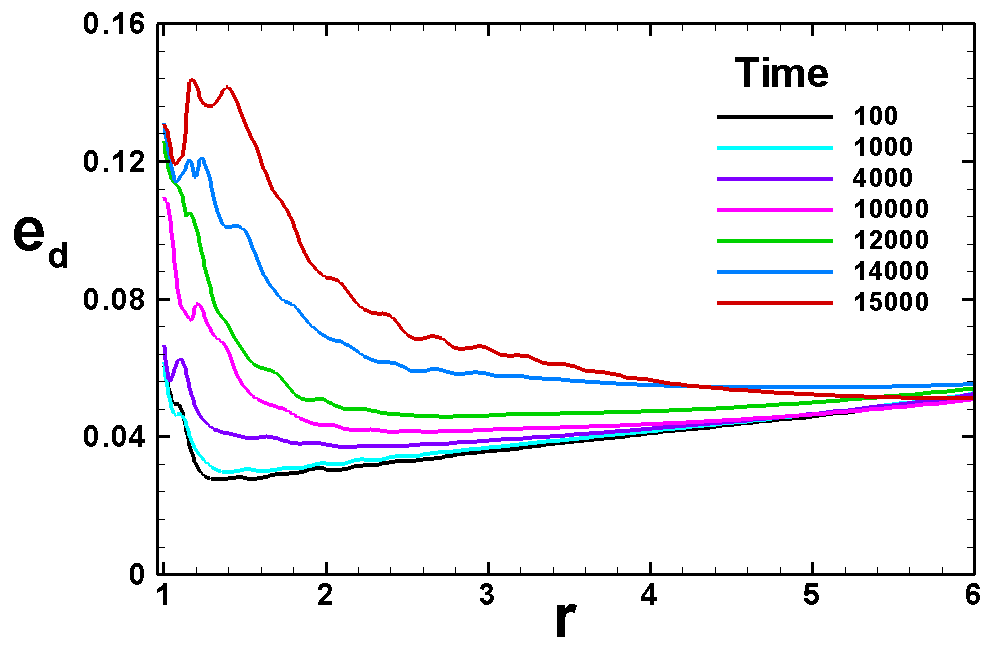} 
     \caption{Distribution of the disc eccentricity $e_d$ with the distance from the star, $r$, at different times.  
     \label{fig:ecc-disc}}
\end{figure}

\subsection{Eccentricity of the disc}
\label{sec:ecc}

The eccentricity of the planet changes due to interaction with the disc, and therefore the eccentricity of the disc also varies with time. 
We observed that the disc becomes more and more non-axisymmetric with time.  Fig. \ref{fig:2d-sigma-8} shows the density distribution at the end of the simulation run in models with different reference disc masses  $q_d$ from the high one, $q_d=1\times10^{-2}$, to the low one,
  $q_d=1\times10^{-4}$.
Top panels of  Fig. \ref{fig:2d-sigma-8}  show
that in the inner parts of the disc, where resonant interaction occurs, the disc is non-axisymmetric. The bottom panels show 
 that the disc overall becomes slightly non-axisymmetric.

We calculate the distribution of the eccentricity of the disc with radius using an approach based on the angular momentum deficit  $A_d(r)$  (e.g., \citealt{RagusaEtAl2018}). For that we calculate the circular angular momentum of the ring in the disc located at radius r:
$$
J_{\rm circ}(r) = \int \Sigma \sqrt{GM a}~d\phi~,~~~~a = - \frac{GM}{2E}~,~~~~E = - \frac{GM}{r} + \frac{v^2}{2}
$$
and the real angular momentum of the ring at the radius $r$:
$$
J_d (r)= \int \Sigma~r v_\phi d\phi .
$$
The angular momentum deficit of the ring is
$$
A_d(r) =J_{\rm circ}(r) - J_d(r)
$$
and the eccentricity of the ring:
$$
e_d(r) = \sqrt {\frac{2~A_d(r)}{J_{\rm circ}(r)} } .
$$

Fig. \ref{fig:ecc-disc} shows the distribution of $e_d(r)$ at different moments in time in the inner part of the simulation region (at larger radii the density of the disc drops due to the exponential cut). One can see that the disc eccentricity increases with time, and it is larger in the inner parts of the disc.   

\citet{RagusaEtAl2018} observed the long-period exchange of eccentricity between the planet and the disc during the stage when the resonant interaction of the planet with the inner disc is small. We suggest that this type of evolution should be further studied in the future.

\section{Discussion}
\label{sec:discussion}

\subsection{Size of the cavity}
\label{sec:size}

The resonant interaction of a planet with the disc is a powerful mechanism for the excitation of  planetary eccentricity.  A planet should be located
 at a distance of $r_p\gtrsim 0.5 r_d$ from the star for the mechanism to operate. The planet's    
eccentricity grows if the size of the cavity does not change significantly with time.  In reality, the size of the cavity  
may vary, for example, due to  variations in the accretion rate. We can compare the time scale of eccentricity growth with the viscous time
scale.

The viscous time scale at the radius $r$  is :
\begin{equation}
         t_{\rm vis}\approx \frac{r^2}{\nu_{\rm vis}}=\frac{r^2}{\alpha c_s H}=\frac{r}{\alpha h^2 v_K},
\end{equation}
where $\nu_{\rm vis}$ is the viscosity coefficient.  
 Normalizing to
our reference values $r_0$, $v_0$ 
and $P_0$, we obtain the dimensionless viscous time scale:
\begin{equation}
\tilde{t}_{\rm vis}=\frac{t_{\rm vis}}{P_0}=\frac{1}{2\pi\alpha h^2}=2.12\times 10^5 \bigg(\frac{3\times 10^{-4}}{\alpha}\bigg)\bigg(\frac{0.05}{h}\bigg)^2 .
\label{eq:tvisc-2}
\end{equation}

\begin{figure*}
     \centering
     \includegraphics[width=0.8\textwidth]{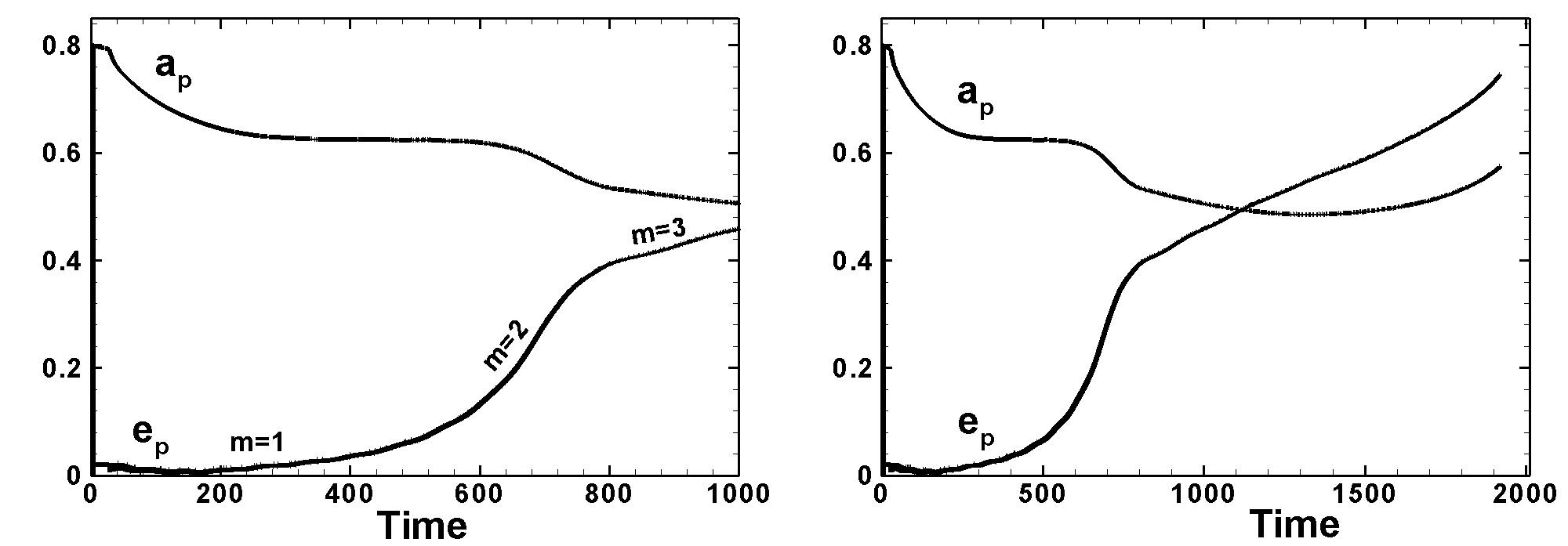} 
     \caption{{\it Left panel:} Temporal evolution of the semi-major axis $a_p$ and eccentricity $e_p$ 
in our model with $M_p=10$ Jupiter mass and other  parameters similar to those of \citet{RiceEtAl2008}. 
 We stopped simulations at $1,000$ rotations to compare this plot with Fig. 7 of R08.  
We observe $m=1, 2,3$ waves in the inner disc, like in our main simulations. \textit{Right panel:} the same, but 
at a longer  simulation time. 
\label{fig:Rice-a08-m10}}
\end{figure*}

\begin{figure*}
     \centering
     \includegraphics[width=0.8\textwidth]{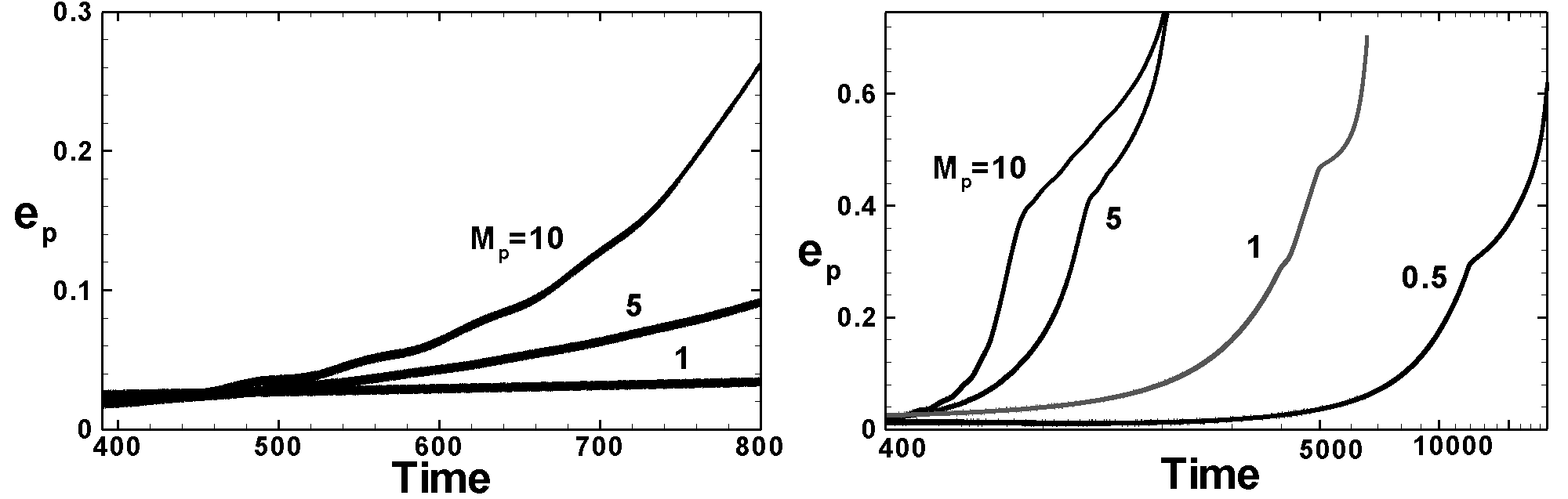} 
     \caption{{\it Left panel:} Eccentricity evolution for planets of different mass $M_p$ and initial $a_p=0.7$.
Simulations are stopped at the time when a planet with $M_p=10$ riched eccentricity $e_p\approx 0.26$ like in Fig. 8
 of R08. The right panel shows that at longer time scales, the eccentricity increased to high values in all models, but the time scale
is much longer for planets of lower mass.   
\label{fig:Rice-a07-mp}}
\end{figure*}

The time scales of the eccentricity growth $\tau_2$ and $\tau_3$ are determined by Eqs.  \ref{eq:phase2} and  \ref{eq:phase3}. We put $t_{\rm vis}=\tau_{2}$  and derive a critical value of $\alpha$ at which the time scales are equal in Phase 2:
\begin{equation}
 \alpha_2=1.62\times 10^{-2} \bigg(\frac{M_p}{10 M_J}\bigg)^{0.91}\bigg(\frac{q_d}{3\times 10^{-4}}\bigg)^{0.86}\bigg(\frac{h}{0.05}\bigg)^{-2.87} .
\label{eq:alpha-2}
\end{equation}
Here, we neglect
 the dependencies on 
$\gamma$ and $n$. If $\alpha$ is smaller than this critical value, then the eccentricity of the planet increases faster than the disc evolves due to viscosity. 
One can see that for the parameters of our model and typical $\alpha=10^{-3}-10^{-4}$, the
eccentricity increases much faster than the viscous disc evolves. We note that  in systems with low-mass planets and/or low-mass discs,
the disc may evolve faster than eccentricity growth. This may explain why more massive planets $\sim (5-30) M_J$ have higher eccentricities
compared to the lower-mass Jovian planets.  The variation of the accretion rate may lead to episodes of eccentricity growth and damping
during protoplanetary disc evolution.

If a planet is located at $r_p<0.5 r_d$, then resonant interaction becomes ineffective. However, the disc and the planet may continue exchanging angular momentum, and  the eccentricities of the planet and disc may increase/decrease in anti-phase \citep{RagusaEtAl2018}. From Fig. 11 of these authors, one can see that this type of evolution 
lasts 10-20 times longer than the time scale of the initial eccentricity growth. In our simulations, we observed that the eccentricity of the disc also increases with time. However, we did not see quasi-periodic variations of eccentricities.   Much longer  simulations are probably required
to study this phenomenon.

Different types of discs and cavities  may have their own specifics  which can affect the planetary eccentricity evolution. For example, if a planet is  located inside the magnetospheric cavity around a young star (e.g.,  classical T Tauri star), then
the size of the magnetosphere is $r_d\sim(3-10) R_*$, and the size of a star should be taken into account because
 a planet on the eccentric orbit may collide with the star \citep{RiceEtAl2008}. On the other hand,
 the tidal interaction of a planet with a star tends to decrease the eccentricity.  Another factor is that a star may accrete in the unstable regime, where tongues of matter penetrate the magnetosphere in the equatorial plane  (e.g. \citealt{KulkarniRomanova2008,KulkarniRomanova2009,RomanovaEtAl2008}). This matter may decrease the  eccentricity of the planet due to the action of the co-orbital corotation torque. This mechanism is more efficient in the cases of relatively small slowly-rotating magnetospheres \citep{BlinovaEtAl2016}. At later stages, the magnetosphere gradually expands and the unstable regime becomes less important.

The low-density cavities may form at different distances from the star and the disc-planet resonant interaction may lead to the formation of eccentric planets at different distances. We have found that the eccentricity increases faster in the case of more massive planets\footnote{Note that a similar result has been obtained by \citet{PapaloizouEtAl2001}.}. This may  
explain the larger eccentricities of more massive planets. 

We should note that if the cavity is very large then the time scale of eccentricity growth may be larger than the lifetime of
protoplanetary disc \citep{DebrasEtAl2021}.    Taking Eq. \ref{eq:phase2} and equating $P_0\tau_2=10^6$ yr (where $P_0=r_0/v_0=r_0^{3/2}/\sqrt{GM_*}$ is our dimensional time scale), we obtain the maximum size of the cavity:
\begin{equation}
r_{\rm max}\approx 51{\rm AU}   \bigg(\frac{T_{\rm disc}}{10^6{\rm yr}}\bigg)^{2/3} \bigg(\frac{M_p}{10 M_J}\bigg)^{2/3}\bigg(\frac{q_d}{3\times 10^{-4}}\bigg)^{2/3} ,
\end{equation}
where $T_{\rm disc}$ is the lifetime of the disc. This formula shows that more massive planets located in more massive discs have enough time to acquire eccentricity even in large-scale cavities.

Our simulations show that the disc eccentricity also increases during the simulations. The angular momentum exchange 
between the planet and the disc may be more complex over longer time scales (e.g.,  \citealt{RagusaEtAl2018,LiLai2022}).
This issue can be studied in the future.

If a planet is located in the large-scale cavity ($\gtrsim10-30$ AU), then inhomogeneities in the disc created by a planet in eccentric orbit could be observed by the ALMA telescope.   
\citet{BaruteauEtAl2021} developed maps in the disc in CO molecular spectral lines which can be compared with ALMA observations. In our models, from 2 to 4 spiral arms are expected in the disc during different stages of eccentricity growth. However, the main feature is the crescent-shaped blob that results from growing eccentricity in the inner disc. This blob may form and dominate the disc structure in particular in models with more massive planets (see the right four panels in Fig. 7; see also \citealt{RagusaEtAl2018}). This issue should be studied in the future.

\begin{figure*}
     \centering
     \includegraphics[width=0.8\textwidth]{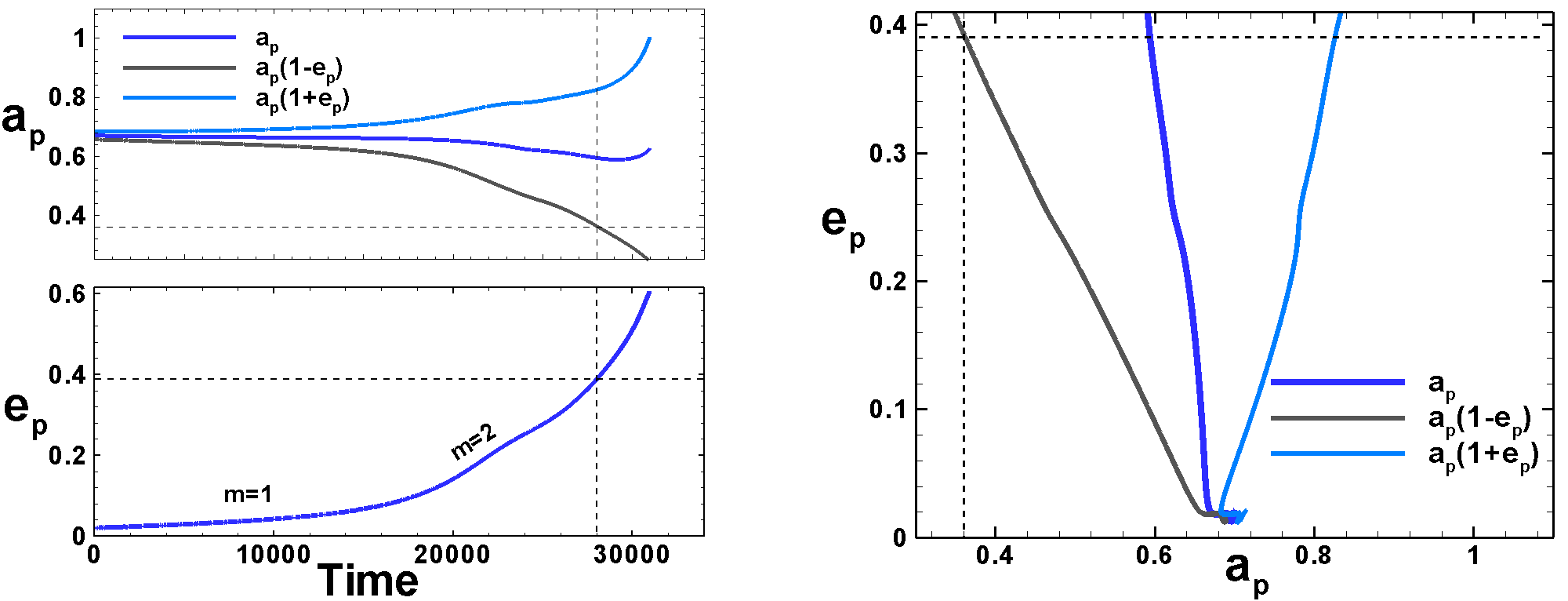} 
     \caption{{\it Left two panels:} Evolution of the semi-major axis $a_p$ and eccentricity $e_p$ in our model with parameters close to those of D21. 
Dashed lines show parameters at which eccentricity stopped increasing in the D21 model due to the inner damping zone region.
\textit{Right panel:} Dependence $e_p-a_p$.  This figure should be compared with Fig. 3 from D21. 
 \label{fig:Debras-a-e}}
\end{figure*}

\subsection{Comparisons with earlier developed models}

\subsubsection{Comparison with model by \citet{RiceEtAl2008}}

Our model is very close to the model of \citet{RiceEtAl2008} (hereafter, R08).  To compare our models, we chose parameters close to those used 
by R08: we placed an empty gap at $r<1$ and considered the simulation domain between $r_{\rm in}=1$ and $r_{\rm out}=10$. The disc has surface density distribution $\Sigma=\Sigma_0  r^{-1}$, where the surface density at $r=1$ is $\Sigma_0=10^{-2}$, which is 30 times larger than the density in our simulations.  We took viscosity with $\alpha=10^{-3}$ in the whole simulation region, while they took $\alpha=10^{-3}$ at $r=1$ and higher values $\sim r$  at larger distances. We used our type of grid with square-shaped grid cells and our typical resolution in $\phi-$direction $N_\phi=640$. The number of grids in the radial direction is $N_r=273$. R08 used grid resolution $400\times400$, which has a lower resolution in the azimuthal direction and compressed grids in the $r-$ direction. They used the code {\it ZEUS} \citep{NormanStone1992}, while we used a Godunov-type code \citep{KoldobaEtAl2016}.

In one experiment, we  took a planet of $M_p=10$ Jupiter mass and placed it initially at $a_p=0.8$. Fig. \ref{fig:Rice-a08-m10} shows the results of simulations. The left panel shows the evolution of the semimajor axis and eccentricity up to time $t=1,000$. We can compare this plot with Fig. 7 of \citet{RiceEtAl2008}. These authors  placed a planet at $a_p=0.9$.  One can see that there is a quantitative and qualitative similarity between these two plots. In both models, the eccentricity increased up to  $e_p\approx 0.4-0.45$.
 In their model, wavy oscillations of eccentricity are often observed, probably due to their boundary conditions. In our model, we used \citet{Val-BorroEtAl2006} boundary conditions, which provide very good wave damping at the boundary, and curves are smoother. \citet{RiceEtAl2008} commented in their paper that they cannot run the code for too long. That is why we continued running the same model up to the moment when a planet in apocenter reached the disc-cavity boundary at $r=1$. The right panel of  Fig. \ref{fig:Rice-a08-m10} shows that the eccentricity reached the value $e_p\approx 0.75$ at the end of simulations. At the end of simulations, the planet reached the value of $a_{\rm ap}\approx 0.2$ in the apocentre. \citet{RiceEtAl2008}  concentrated on the problem of planet survival in magnetospheres of young stars, where the typical size of the magnetosphere is $r_m\sim 3-10$ stellar radii. The authors concluded that the high-mass planets may not survive in such cavities due to the high eccentricity and collision with the star. Though this conclusion is confirmed by our simulations, we note that in cavities of larger size, the planet eccentricity may increase up to high values. We checked the types of resonances operating at these parameters. 2D plots of density distribution in the disc show the presence of waves with $m=1, 2, 3$ spiral arms.

In another experiment, we took the initial position of the planet at $a_p=0.7$ and calculated the eccentricity evolution at different masses of the planet, taken by R08: $M_p=10, 5, 1, 0.5$.  The left panel of Fig. \ref{fig:Rice-a07-mp} shows eccentricity evolution during a relatively brief time interval. This plot can be compared with Fig. 8 of R08. These figures show qualitatively similar results: eccentricity increases fast in the model with $M_p=10$ Jupiter mass and slower in the model with  $M_p=5$. At masses $M_p=1$ and $0.5$, the eccentricity seems to be not growing \footnote{The time scales of eccentricity growth do not exactly coincide due to a few differences in models, e.g., different boundary conditions, different energy equations (they used locally-isothermal disc), and more.} We continued running these models for a longer time and obtain that in all models, the eccentricity increases up to a high value of $e_p\approx 0.7-0.75$, though the time scale strongly increases for planets of lower mass. The right panel of Fig. \ref{fig:Rice-a07-mp} shows that the eccentricity reached high values in all models. We conclude that there is no theoretical obstacle to the eccentricity growth of a lower mass planet, but it takes much longer time, which in some cases may be unrealistically long (see Sec. \ref{sec:size}).

\subsubsection{Comparison with model by \citet{DebrasEtAl2021}}

We also compared the results of our model with results obtained by \citet{DebrasEtAl2021} (hereafter, D21) who performed
 simulations both in the disc and in the cavity and obtained the maximum eccentricity $e_p\lesssim 0.4$. They supported a low-density cavity by placing a low viscosity in the disc and high viscosity in the cavity and were able to obtain a quasi-stationary cavity at the radius $r=1$. The authors had a goal to investigate  the eccentricity growth of planets of Jupiter's mass.

We took our model but chose parameters close to those of D21: mass of the planet: $M_p=1$ Jupiter mass, reference density in the disc is 3 times higher than in our model: $\Sigma_0=10^{-3}$. We took a flat density distribution with $n=l=0$, like in D21. 
We took a small simulation region with the outer radius $r_{\rm out}=2$, like in D21 (who had $r_{\rm out}=1.9$). We used our typical grid resolution $N_\phi=640$  in the $\phi$ direction,
and grid $N_r=84$ in the radial direction (which resulted from our preference to keep square-shaped grid cells). We note that D21 used mainly  grids $200\times 200$ or $400\times 400$. Also, D21 used a locally-isothermal equation of state (with the fixed temperature distribution), while we solved the energy equation and took $\gamma=5/3$.

Despite a few differences,  our model is in reasonably good agreement with that of  D21.  Left panels of  Fig.  \ref{fig:Debras-a-e} show
that in our model the eccentricity increased up to the value of $e_p\approx 0.6$ during $\approx 30,000$ rotations of the inner disc. Our simulations were stopped when a planet in the apocentre reached $R_{\rm ap}=1$. The right panel shows the dependence of the eccentricity on the semi-major axis in the style of Fig. 3 of D21. Comparisons of this plot with the right panel of Fig. 3 of D01 show that in D21 the eccentricity stops growing at $e_p\approx 0.35-0.4$. The authors explained that a planet enters a wave-damping zone
placed at  $r=0.36$. In our model, the cavity is empty and eccentricity increases to larger values. However, if we take a part of our curves 
(restricting the apocenter of the planet with the value  of
 $a_{\rm per}\approx 0.36$) then we obtain the maximum eccentricity of $e_p\approx 0.4$ which corresponds to that obtained in the D21 model. We note that in this interval of parameters, our curves are similar.
Our time scale 
of eccentricity growth up to the value $e_p=0.4$ is $28,000$ rotations which is close to $20,000$ rotations obtained by D21 at the grid $400\times 400$ (see their Fig. 3). 
We note that in D21, the time scale increases up to $60,000$ rotations at their runs with the lower grid resolution, $200\times 200$.  Overall, our result is close to the result of D21 at their higher grid resolution and at parameters $a_{\rm ap}\gtrsim 0.0.36$ and $e_p\lesssim 0.4$.

We also plotted the 2D slices of the surface density distribution and observed the formation of the $m=1$ one-armed spiral wave in the disc
at times $\lesssim 15,000$, and $m=2$ two-armed spiral waves (corresponding to the 1:3 ELR) during the rest of simulation time. We observed that waves were jammed in the small-sized simulation region of $1<r<2$. This may be the reason why $m=3$ waves  (the $2:4$ resonance) were not observed.

\subsubsection{Comparison with model by \citet{PapaloizouEtAl2001}}

\citet{PapaloizouEtAl2001} (hereafter P01) investigated the migration  of planets with masses $M_p=1-30$ Jupiter mass 
in the disc, which gradually opened a central cavity and subsequently interacted with the disc. They observed that the planet's 
eccentricity increases due to interaction with the disc. It increases more rapidly in models with higher mass $M_p$ (see their Fig. 1). 
This is in accord with our models  (see Sec. \ref{sec:mass} and our Fig. \ref{fig:orb-dedt-mass}).

P01 also analyzed the causes of eccentricity growth.  They fixed a $30 M_J$ mass planet at the orbit and analyzed the role of 1:2 and 1:3 resonances in the eccentricity growth. They observed that the action of the 1:2 resonance is slightly  stronger than that of the 1:3 resonance and concluded that the primary cause of the eccentricity growth is the excitation of the eccentricity in the disc and subsequent interchange of angular momentum between the disc and the planet. We note here that when the planet is fixed (and eccentricity is zero) then the action of the 1:2 ECR is larger than 1:3 ELR and corotation torque suppresses the eccentricity growth (e.g., \citealt{GoldreichSari2003,OgilvieLubow2003}). We observed it in our simulations which were initially performed at $e_p=0$.  Later, we switched to models with initial $e_p=0.02$ (like, e.g.,  R08 did).  The right panel of our Fig. 1 shows the difference between models with $e_p=0$ and $0.02$.

P01 noted that the eccentricity does not increase in their model with $M_p=1$ Jupiter mass. 
 In our models, we observed that eccentricity of  $ 1 M_J$ planet increases to a high value of $e_p\approx 0.6$ but very slowly compared with 
more massive planets (see our Fig. \ref{fig:orb-dedt-mass} where the initial part of the eccentricity growth is shown). 

P01 commented that the eccentricity of a Jupiter mass planet  may start growing at the lower viscosity of the disc.
Our simulations performed at different values of viscosity parameter $\alpha$ show that viscosity damps resonant waves and  decreases  the rate of eccentricity growth. 
We also note that modeling a Jupiter mass planet requires a higher grid resolution, compared with more massive planets.

\subsubsection{Comparison with model by  \citet{RagusaEtAl2018}}

\citet{RagusaEtAl2018} (hereafter R18) performed simulations of two models with the mass of the planet $M_p=13$ Jupiter mass and
two discs with the reference surface densities $\Sigma_0=1.5\times 10^{-4}$ and $\Sigma_0=4.8\times 10^{-5}$. Simulations were performed
at a high grid resolution $430\times 580$ and in a large simulation region: $0.2<r<15$ with an additional exponential taper at $r=5$.
They observed the formation of the crescent-shaped overdense feature at the apocentre of the cavity which is consistent with the density perturbation expected for an eccentric disc \citep{TeyssandierOgilvie2016}. Our simulations also show the formation of such a feature which is particularly clear in models 
with a high planet mass (see right panels in our Fig. \ref{fig:2d-mass-8}). We expect that this feature can explain some of the ALMA observations where the crescent-shaped brightness enhancement is often observed (see also \citealt{AtaieeEtAl2013,RagusaEtAl2017}). This is a similarity between our models. 

In models of  R18, a planet eccentricity does not increase above a small value of $r_p\approx 0.14$.  Our models with comparably low density of $\Sigma=q_d=10^{-4}$ show that the planet eccentricity increases up to high values,  but very slowly (see our Fig. \ref{fig:orb-dedt-sigma}).   It is not clear why
eccentricity does not increase to larger values in models of R08. 
The authors do not show details of early times when the planet could excite resonances
in the disc. In their Fig. 1, the inner disc is located at the distance of $\sim (2-3) a_p$ at which all resonances are inside the cavity.  
In this work,  the stage of resonant interaction may have been relatively brief and eccentricity increased only a part-way. 
It is possible that in R18 the inner disc moved outward faster than the rate of eccentricity growth. 

At later times, when the inner disc is too far away and the  resonances do  not operate, R18 calculated the interaction of a planet with the disc and found interesting long-term variability where the planet and the disc exchange their angular momentum and their eccentricity vary in the anti-phase. 
 These types of simulations are beyond the scope of our paper.

\section{Conclusions} \label{sec:conclusions}

We have investigated the growth of the eccentricity of massive planets located 
inside the cavities of protoplanetary discs  due to the resonant planet-disc interactions. 
 The main conclusions  are the following:

\smallskip

\noindent \textbf{1.}  We observed clear, long-lasting resonant interactions between the disc and the planet driven by different
resonances. This helped us to investigate the properties of such interaction in detail.

\smallskip

\noindent \textbf{2.}  In most of the simulations, the planet's  eccentricity grows to a large value of $e_p\approx 0.65-0.75$, which have 
never been obtained in previous  numerical works.  
Note that we stopped our simulations  when the planet in eccentric orbit reaches the cavity boundary. Otherwise, the eccentricity could increase to even larger values.

\smallskip

\noindent \textbf{3.} The planet's eccentricity growth proceeds through several distinct phases: 
(1) A slow exponential growth due to the 1:2 OLR at which one-armed  (m=1) spiral waves are excited in the disc. (2) Rapid  growth due to the 1:3 (m=2) ELR up to $e\approx 0.2-0.25$. (3) Slower growth up to $e\approx 0.5$ due to the 2:4 (m=3) ELR resonance.  (4) A relatively brief time
 interval  of eccentricity growth up to $e\approx 0.65-0.75$, in which the
$m=4$ waves are observed. This phase may correspond to the excitation of the 3:5 resonance.

\smallskip

\noindent\textbf{4.} We  varied the mass of the planet and various parameters of the disc in order to derive
 the dependencies of the 
eccentricity growth rate on these parameters. The growth rate driven by the 1:3 ELR is proportional to the
 planet's mass and the disc surface density (for a wide interval of parameters), in agreement with  theoretical predictions. The growth rate decreases with the $\alpha$-parameter of viscosity and the thickness of the disc $h$. In Phase 3, the eccentricity grows 2-3 times slower. 
Many other dependencies are complex and are presented as figures 
 (see Figs. \ref{fig:orb-dedt-sigma}-\ref{fig:orb-dedt-slope}). In the vicinity of reference values, we derived 
analytical dependencies  (see Eqs. \ref{eq:phase2} and \ref{eq:phase3}).

\smallskip

\noindent\textbf{5.}  We derived the width of the 1:3 ELR from numerical simulations:  $w_L\approx 0.19 r_d$ which
is  $\sim 2.3$ times larger than that predicted by the theory (e.g., \citealt{TeyssandierOgilvie2016}).
We observed that the resonance may be inside the cavity but the planet interacts with the part of the resonance due to its finite width. 

\smallskip

\noindent\textbf{6.} The eccentricity of the planet can  grow if the time scale of the  growth is shorter than that of the cavity evolution due to viscosity. For a wide range of parameters, the eccentricity growth time  is indeed shorter than the viscous time scale
(see Eq. \ref{eq:alpha-2}).

\smallskip

\noindent\textbf{7.} The results obtained in our numerical models may help in understanding the non-linear stages of eccentricity growth  
and the development of the theory of non-linear resonant interaction.  

\smallskip

A caveat of the simulations reported in this paper is that  we fixed the boundary of the cavity, inside which the disc matter does not penetrate. This helped us to exclude the local corotation torque, which damps the planet's eccentricity.   This setup also helped us to investigate the role of different resonances in the eccentricity growth and the dependencies of the eccentricity growth rate on various  physical parameters. As a next step, we plan to investigate resonances and planetary eccentricity growth in 2D and 3D simulations where the cavity has low density. The knowledge obtained in the current research will help us to choose the  parameters of future simulations.

\section*{Acknowledgments}
Resources supporting this work were provided by the NASA High-End
Computing (HEC) Program through the NASA Advanced Supercomputing
(NAS) Division at Ames Research Center and the NASA Center for
Computational Sciences (NCCS) at Goddard Space Flight Center. MMR and RVEL were supported
in part by the NSF grant AST-2009820. The authors thank Karan Baijal for editing the manuscript and the
anonymous referee for important comments, questions, and suggestions.

\section{Data Availability}

The data underlying this article will be shared on reasonable request to the corresponding author (MMR).

\bibliographystyle{mn2e}

\appendix

\section{Details of the numerical model}
\label{sec:numerical}

We calculate the evolution of the disc and the orbit of the planet in the coordinate system 
centered on the star. This coordinate system is not inertial due to the presence of the planet and the disc \footnote{In our models, the integrated force from the disc onto the star is a few orders of magnitude smaller than that from the planet, and we neglect the inertial term associated with the disc.}.
 That is why in equations of motion 
for the disc and the planet, we add an additional term for the  inertial force.
We solve the hydrodynamic equations in polar coordinates ($r,\phi$):   

        \begin{equation}
                   \nonumber  \frac{\partial\Sigma}{\partial t}
                            + \frac{1}{r}\frac{\partial}{\partial r}(r\Sigma v_{r})
                            + \frac{1}{r}\frac{\partial}{\partial\varphi}(\Sigma v_{\varphi}) = 0,
        \label{eqn:continuity}
        \end{equation}

    \begin{eqnarray}
        \nonumber \frac{\partial}{\partial t}(\Sigma v_{r})
                                        &+& \frac{1}{r}\frac{\partial}{\partial r}\left[r\left(\Sigma v_{r}^{2}
                                       + \Pi\right)\right] + \frac{1}{r}\frac{\partial}{\partial\varphi}\left(\Sigma v_{r}v_{\varphi}\right) \\
                                        &=& \frac{\Pi}{r} - \Sigma\frac{GM_{\star}}{r^{2}} + \Sigma w_{r}
    \label{eqn:radial_motion}
    \end{eqnarray}
        \begin{eqnarray}
            \nonumber \frac{\partial}{\partial t}(\Sigma v_{\varphi})
                                &+& \frac{1}{r^{2}}\frac{\partial}{\partial r}
            \left[r^{2}\left(\Sigma v_{r}v_{\varphi}\right)\right]  + \frac{1}{r}\frac{\partial}{\partial\varphi}
                                \left(\Sigma v_{\varphi}^{2} + \Pi \right) = \Sigma w_{\varphi}
            \label{eqn:az_motion}
        \end{eqnarray}
        \begin{equation}
                 \nonumber    \frac{\partial}{\partial t}\left(\Sigma S\right)
               + \frac{1}{r}\frac{\partial}{\partial r}\left(r\Sigma S v_{r}\right)
                            + \frac{1}{r}\frac{\partial}{\partial\varphi}(\Sigma S v_{\varphi}) = 0.
        \end{equation}

  \noindent Here $\Sigma=\int\rho dz$ is the surface density (with $\rho$ the volume density);  $v_{r}$ and $v_{\varphi}$ are the radial and azimuthal
    velocities, respectively;
    $\Pi = \int P dz$ is the surface pressure (with $P$ the volume pressure); $S = {\Pi}/{\Sigma^{\gamma}}$ is a function analogous to entropy; and  $\gamma$ is adiabatic index.  $w_{r}$ and 
$w_{\varphi}$ are the forces exerted on the disc by the planet (per unit area of the disc).

\begin{figure*}
     \centering
     \includegraphics[height=0.27\textwidth]{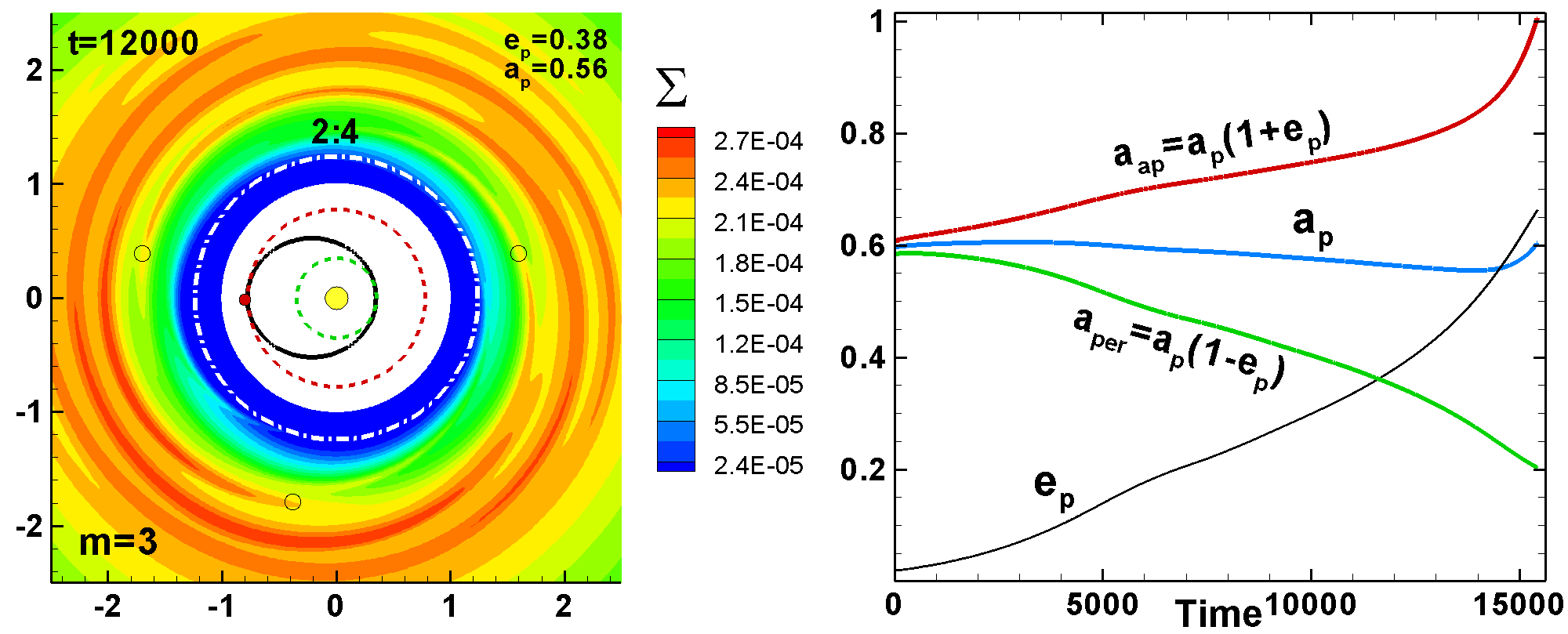}
     \includegraphics[height=0.27\textwidth]{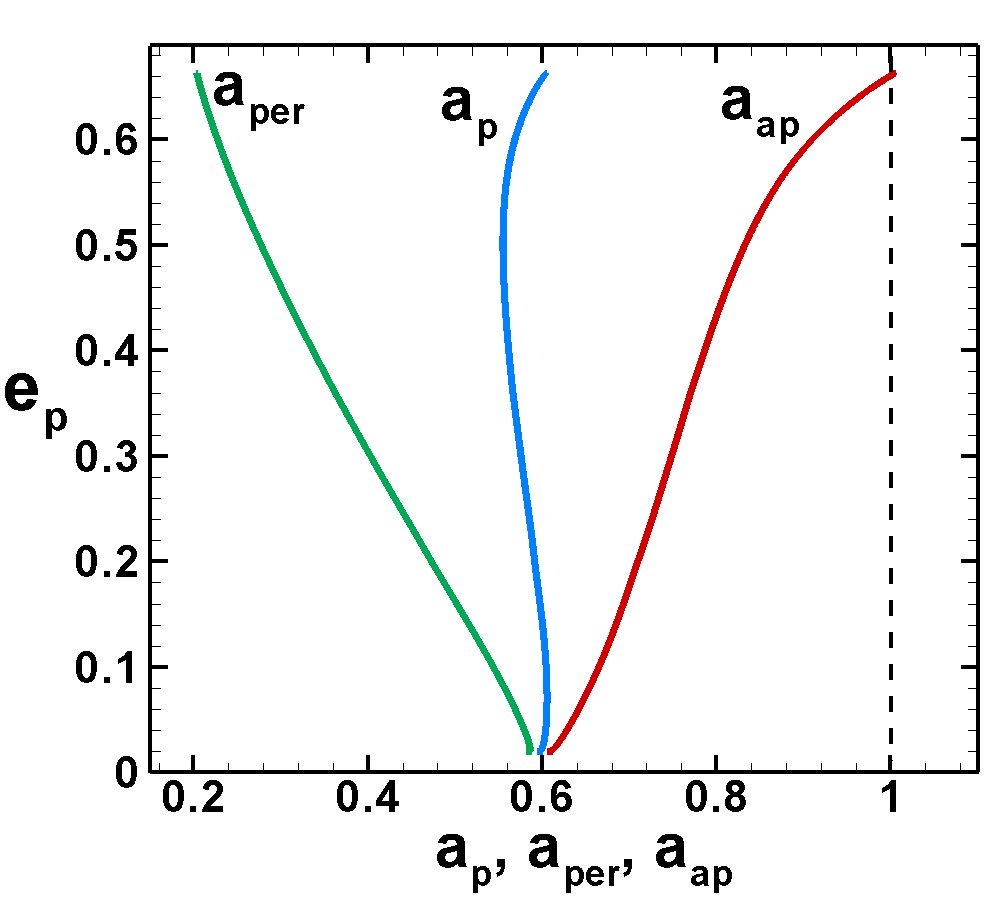} 
     \caption{\textit{Left Panel:} Sketch of the main resonances plotted on top of the density distribution in the reference model with $a_p=0.6$ 
at $t=12,000$ when 2:4 resonance with $m=3$ dominates. Resonances were calculated using the apocentre  $a_{\rm ap}=a_p(1+e_p)$ position, marked as an outer dashed circle (instead of the semimajor axis $a_p$). \textit{Middle Panel:} Temporal variation of the semi-major axis $a_p$, apocentre $r_{\rm ap}$  and pericentre $r_{\rm per}=a_p(1-e_p) $ in the reference model. The black line shows eccentricity. \textit{Right panel:}  Dependence of the planet's eccentricity $e_p$ on the semimajor axis $a_p$, apocentre $a_{\rm ap}$
 and pericentre $a_{\rm per}$ . 
     \label{fig:sketch-2}}
\end{figure*}

\begin{figure*}
     \centering
     \includegraphics[width=0.8\textwidth]{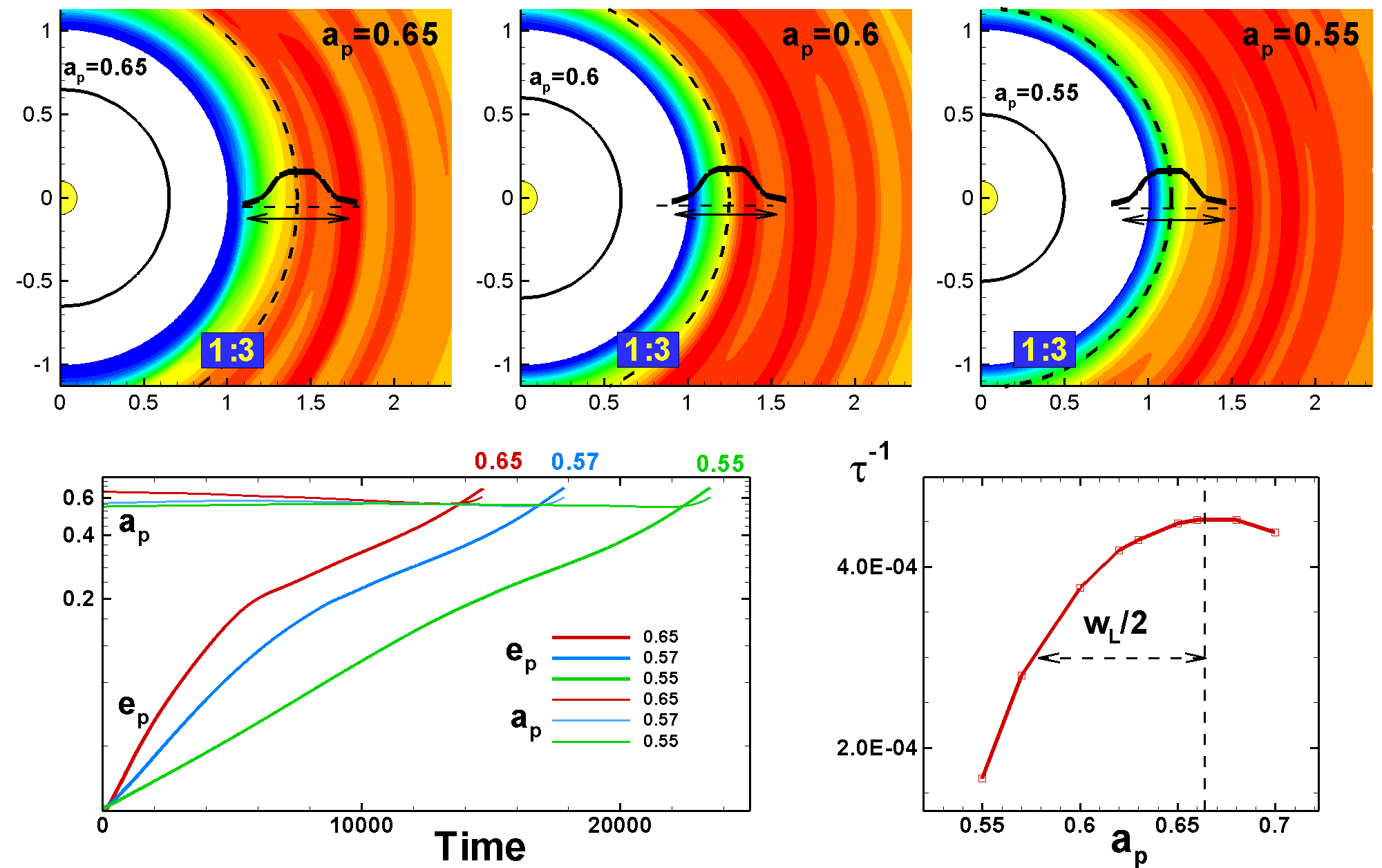} 
     \caption{{\it Top panels:} Surface density distribution (color background), initial positions of the planet (solid lines),
and position of the 1:3 ELR (dashed bold line).  The approximate width of the ELR is shown with arrows and
schematics.   
{\it Bottom panels:} {\it Left:} Temporal variation of  eccentricity in models with different $a_p$. {\it Right:} Dependence of the eccentricity growth rate $\tau^{-1}$ in Phase 2 on the initial semi-major axis of the planet, $a_p$.  The horizontal line with arrows shows the half-width of the ELR.  
\label{fig:2d-sax-5}}
\end{figure*}

Viscosity terms are added to the equations of motion
following the $\alpha$ prescription of \citet{ShakuraSunyaev1973}, 
with the  viscosity coefficient in
the form of $\alpha-$viscosity, $\nu_{\rm vis}=\alpha c_s H$.

In our code, we use the entropy balance equation instead of the full
energy equation, because in the problems that we solve, the shock
waves (where we cannot neglect the energy dissipation) are not expected.
This approach is more appropriate for the investigation of waves in the disc compared 
with the widely used locally-isothermal approach, where the
temperature is fixed in time and depends on the radius only (e.g., \citealt{RagusaEtAl2018,DebrasEtAl2021};
see also discussion of this issue in \citealt{MirandaRafikov2020}).

The equations of hydrodynamics are integrated numerically
using an explicit conservative Godunov-type numerical scheme
\citep{KoldobaEtAl2016}.  
For the calculation of fluxes between the cells, we use the HLLD Riemann's
solver developed by \citet{MiyoshiKusano2005}.
 Integration of the
equations with time are performed with a two-step Runge-Kutta
method. 

Our code is similar in many respects to other codes which use
Godunov-type method, such as PLUTO (Mignone et al. 2007),
FLASH (Fryxell et al. 2000), and ATHENA (Stone et al. 2008). 
The code is different from FARGO3D code where the orbital advection has been implemented \citep{Bentez-LlambayMasset2016}.
Our code has been thoroughly tested using standard tests \citep{KoldobaEtAl2016} 
\footnote{In addition, it has been tested on several astrophysical problems. In 2D hydro and MHD versions, it has
been used for modeling planet migration in accretion disc \citep{CominsEtAl2016}. In the case of a non-magnetized disc with different slopes in density distribution, we obtain the transition from inward to outward migration at the slope which is close to that predicted theoretically by \citet{Tanaka2002}. In the model of a magnetized disc, we obtained the positions of
the magnetic resonances at locations similar to those found in the simulations
by \citet{FromangEtAl2005}, which both correspond to the theoretical resonance locations predicted by \citet{Terquem2003}. A 3D hydro version of the code has been used to study the trapping of low-mass planets
 at the disc-cavity boundary due to the corotation torque \citep{RomanovaEtAl2018}. In models with thin discs, our results have shown trapping radii similar to those obtained in 2D simulations by \citet{MassetEtAl2006}.}. 

The code is parallelized using MPI. We typically use 448 processors and run the code during 20-200 hours, depending on the parameters.
Simulations are longer in models with lower disc mass and smaller masses of the planet.

\smallskip

We calculate the orbit of the planet
using earlier developed approaches (e.g. \citealt{Kley1998,Masset2000,KleyNelson2012}).
The force per unit mass acting on the disc is:
\begin{equation}
\mathbf{f} = -\frac{GM_*}{|\mathbf{r}|^3}\mathbf{r} -
\frac{GM_p}{|\mathbf{r}-\mathbf{r}_p|^3}
\mathbf{(r-r_p)} - \frac{GM_p}{|\mathbf{r}_p|^3} \mathbf{r}_p  ~,
\label{eq:grav_accel}
\end{equation}
where $\mathbf{r}_{\rm p}$ is the radius vector from the star to
the planet. The first and second terms represent the gravitational forces from
the star and the planet respectively. The last term accounts for
the fact that the coordinate system is not inertial.

We find the position $\mathbf{r}_{\rm p}$ and velocity $\mathbf{v}_{\rm p}$ of the planet at each time step solving
the equation of motion:
\begin{eqnarray}
 M_{p} \frac{d\mathbf{v}_{\rm p}}{dt} = -\frac{GM_{\star}M_{\rm p}}{|\mathbf{r}_p|^{3}}\mathbf{r}_p
                                                                              -\frac{GM_{\rm p}^{2}}{|\mathbf{r}_p|^{3}}\mathbf{r}_ p + \mathbf{F}_{\rm disc\rightarrow p},
\label{eq:planet}
\end{eqnarray}
where 
\begin{equation} 
\mathbf{F}_{\rm disc\rightarrow p}=\int\frac{GM_p}{|\mathbf{r}-\mathbf{r}_p|^3}(\mathbf{r}-\mathbf{r}_p)\Sigma r dr d\phi
\end{equation} 
is a cumulative force acting from the disc to the planet. If we neglect the force 
$\mathbf{F}_{\rm disc\rightarrow p}$ in Eq. \ref{eq:planet}, then the trajectory  of the planet is described 
by the equations of motion in the gravitational field of the cumulative mass $M=M_*+M_p$.

We  calculate the planet's orbital energy and angular momentum
per unit mass using the calculated values of $\mathbf{r}_{\rm p}$
and $\mathbf{v}_{\rm p}$:
\begin{align}
    E_p = \frac{1}{2}\lvert{\mathbf v}_{\rm p}\rvert^{2} - \frac{GM}{r_p}
    & & {\rm and}
    & &  L_p = \mathbf{r}_{\rm p} \times \mathbf{v}_{\rm p} .
\end{align}
We use these relationships to calculate the semimajor axis and eccentricity of the planet's orbit at each time step:
\begin{align}
    a_p = -\frac{1}{2}\frac{GM}{E_p} & & {\rm and}
    & &
    e_p = \sqrt{1 - \frac{L_p^{2}}{GM a_p}} .
\end{align}

\section{Effects of the elliptical orbit}
\label{sec:elliptical}

In our reference models, we observe spiral density waves with $m=1, 2, 3$, and $4$ arms which can be excited at the 1:2 OLR and  1:3, 2:4, and 3:5 ELRs. 
 However, higher-order resonances are located closer to the star (see Tab. \ref{tab:resonances}). In reference model with $a_p=0.6$
 and for typical values of $a_p=0.55-0.6$,
many resonances are located inside the cavity. 
 The question arises, why do we see resonant interaction?

On one hand, resonances have a finite width and part of the resonance can be in the disc, as described in Sec. \ref{sec:width}.
This factor should play a role. On the other hand, 
a planet in an elliptical orbit has the closest approach to the disc during its passage through the apocentre, which is located at a distance of
$r_{\rm ap}=a_p (1+e_p)$ from the star.  A planet spends a significant time at this part of the orbit and may excite ELRs 
during its passage through the apocentre. We can calculate the location of resonances using $r_{\rm ap}$ (instead of $a_p$).

We show examples using the reference model with $a_p=0.6$. The right panel of Fig. \ref{fig:sketch-2} shows the temporal evolution of $a_p$, $e_p$, apocentre $r_{\rm ap}$ and pericentre
$r_{\rm per}=a_p (1-e_p)$. 
 One can see that the apocentre increases with time up to 
$r_{\rm ap}=1$ when the planet reaches the disc inner boundary.

For example, at $t=12,000$, the parameters of the orbit are $a_p\approx 0.56$ and $e_p\approx 0.38$.
The left panel of Fig. \ref{fig:sketch-2} shows that the three-armed density wave dominates, which favors
the 2:4 ELR resonance. Using $a_p$ as a base, we obtain  $r_{\rm ELR,2:4}\approx 1.587  r_{\rm ap}\approx 0.89$, 
which is within the cavity. 
The apocentre is located at $r_{\rm ap}\approx 0.78$ and the corresponding 2:4 resonance 
 at 
$r_{\rm ELR,2:4}\approx 1.587 r_{\rm ap}\approx 1.24$, that is, within the disc. This resonance is shown as
the white dashed-dot line in Fig.  \ref{fig:sketch-2}.

Similarly, at  time $t=15,000$ when $m=4$ waves were observed, we calculate  the position of the 3:5 resonance. 
The parameters of the orbit are $a_p=0.57$ and $e_p=0.61$. The position of the planet in the apocentre is 
$r_{\rm ap}\approx 0.92$, and the location of the resonance is at $r_{\rm ELR,3:5}\approx 1.406 r_{\rm ap}\approx 1.29$, which is within the disc.
Note that if we use $a_p$ as a base, we obtain  $r_{\rm ELR,3:5}\approx 0.80$.

\section{The width of the 1:3 ELR resonance}
\label{sec:width}

Resonances in the disc have a finite width which increases with the disc thickness h (e.g., \citealt{GoldreichTremaine1978}). 
In the linear approximation,
 the width has been derived 
from theoretical studies (see e.g., Eq. \ref{eq:width}).  Using
parameters of our reference models $a_p=0.6$, $h=0.05$ and taking $m=2$ for 1:3 ELR, we  obtain  
theoretically predicted width:
\begin{equation} 
w_L\approx 0.065 r_{\rm ELR,1:3}h^{2/3}\approx 0.081\bigg(\frac{h}{0.05}\bigg)^{2/3} ,
\end{equation}
where we take into account that $r_{\rm ELR,1:3}\approx 2.08 a_p$.

Here, we estimate the width of the 1:3 ELR using our numerical model. For that, we 
place a planet at different  distances $a_p$ from the star, from $a_p=0.7$ to $a_p=0.55$, and let them migrate.  The top panels of Fig. \ref{fig:2d-sax-5} 
show the initial orbit of the planet with semimajor axis $a_p$ (solid line) and positions of the 1:3 ELR which are located at $r_{\rm ELR,1:3}\approx 2.08 a_p$ (see bold dashed lines).  One can see that at $a_p=0.65$ (left top panel), the ELR is located at $r_{\rm ELR,1:3}\approx 1.35$, and a significant part of the resonance width is located in the denser part of the disc (red and yellow colors). At $a_p=0.6$ (middle top panel), the ELR is located at $r_{\rm ELR,1:3}\approx 1.25$, and approximately half of the resonance width is located in the disc. At $a_p=0.55$, the resonance is located at $r_{\rm ELR,1:3}\approx 1.14$, and only a part of the resonance width is inside the disc. It is expected that in the model with $a_p=0.65$ the resonant interaction will be stronger than in models with smaller initial $a_p$.

The bottom left panel shows that at larger values of $a_p$, the eccentricity increases faster, as expected. In models with smaller $a_p$ the eccentricity increases, but slower, because only a part of the resonance width is located in the disc.   We  calculated the eccentricity growth rate  $\tau^{-1}=(1/e_p)de_p/dt$ in Phase 2 for models shown in the left panel. The bottom right panel shows that  $\tau^{-1}$ is approximately the same in models with $a_p\gtrsim 0.65$
because all (or most) of the resonant width is inside the disc. However, $\tau^{-1}$ decreases when $a_p$ decreases because a smaller and smaller part of the resonant width is inside the disc. From the curve of Fig. \ref{fig:2d-sax-5}  we can estimate the width of the resonance at half of the amplitude. 
We obtain  $w_L/2\approx 0.08$ and the full width of the resonance is $w_L\approx 0.16$. This value is $\sim 2$ times larger than that derived theoretically.

The finite width of resonances is important in the disc-planet interaction because a planet may interact with the disc 
even if the center of resonance is located in the cavity.

\end{document}